\documentclass{amsart}
\usepackage{enumerate}
\usepackage[dvips]{graphics}
\usepackage[dvips]{graphicx}
\usepackage{epsfig,enumerate,mathrsfs}
\usepackage{color}

\numberwithin{equation}{section}
\newtheorem{theorem}{Theorem}[section]
\newtheorem{proposition}[theorem]{Proposition}
\newtheorem{lemma}[theorem]{Lemma}
\newtheorem{corollary}[theorem]{Corollary}
\newtheorem{remark}[theorem]{Remark}

\newcommand{\la}{\lambda}
\newcommand{\Aa}{\mathcal{A}}
\newcommand{\Bb}{\mathcal{B}}

\newcommand{\z}{\zeta}

\newcommand{\Ll}{\mathcal{L}}

\newcommand{\R}{\mathbb R}
\newcommand{\Z}{\mathbb Z}

\newcommand{\Nn}{\mathcal{N}}

\newcommand{\zd}{{\det}_{\zeta}}
\def\N{\mathbb{N}}
\def\C{\mathbb{C}}
 
\def\Id{\mathrm{Id}}

\def\KD3{\mathrm{KD}^3}
\def\dD{\mathscr{D}}

\def\max{\mathrm{max}}
\def\Ei{\mathrm{Ei}}

\begin{document}

\title[Functional determinants on the generalized cone]
{Functional determinants for general self-adjoint extensions of
Laplace-type operators resulting from the generalized cone}

\author{Klaus Kirsten}
\address{Department of Mathematics\\ Baylor University\\
         Waco\\ TX 76798\\ U.S.A. }
\email{Klaus$\_$Kirsten@baylor.edu}

\author{Paul Loya}
\address{Department of Mathematics \\Binghamton University\\
Vestal Parkway East\\Binghamton\\NY 13902\\ U.S.A. }
\email{paul@math.binghamton.edu}

\author{Jinsung Park}
\address{School of Mathematics\\ Korea Institute for Advanced Study\\
207-43\\ Cheongnyangni 2-dong\\ Dongdaemun-gu\\ Seoul 130-722\\
Korea } \email{jinsung@kias.re.kr}



\begin{abstract}
In this article we consider the zeta regularized determinant of
Laplace-type operators on the generalized cone. For {\it
arbitrary} self-adjoint extensions of a matrix of singular
ordinary differential operators modelled on the generalized cone,
a closed expression for the determinant is given. The result
involves a determinant of an endomorphism of a finite-dimensional
vector space, the endomorphism encoding the self-adjoint extension
chosen. For particular examples, like the Friedrich's extension,
the answer is easily extracted from the general result. In
combination with \cite{BKD}, a closed expression for the
determinant of an arbitrary self-adjoint extension of the full
Laplace-type operator on the generalized cone can be obtained.
\end{abstract}

\maketitle


\section{Introduction}
Motivated by endeavors to give answers to some fundamental
questions in quantum field theory there has been significant
interest in the problem of calculating the determinants of second
order Laplace-type elliptic differential operators; see for
example \cite{birrell,feyn,raja,ramond,schul}. In case the
operator $\Delta$ in question has regular coefficients and is
acting on sections of a vector bundle over a smooth compact
manifold, it will have a discrete eigenvalue spectrum
$\lambda_1\leq \lambda_2 \leq ... \to \infty$. If all eigenvalues
are different from zero the determinant, formally defined by $\det
\Delta = \prod_i \lambda_i$, is generally divergent. In order to
make sense out of it different procedures like Pauli-Villars
regularization \cite{PaVi} or dimensional regularization
\cite{tHooft} have been invented. Mathematically the probably most
pleasing regularization is the zeta function prescription
introduced by Ray and Singer \cite{RS71} (see also
\cite{cri,HawS77}) in the context of analytic torsion; see i.e.
\cite{BGSI88,BGSII88,BGSIII88,Mu78,MuMu}.

In this method, one uses the zeta function $\zeta (s,\Delta )$
associated with the spectrum $\lambda_i$ of $\Delta$. In detail,
for the real part of $s$ large enough one has $$\zeta (s,\Delta )
= \sum_{i=1}^\infty \lambda_i^{-s}.$$ In the briefly described
smooth setting, one can show that $\zeta (s,\Delta )$ is analytic
about $s=0$ \cite{BGiP95,SeR67,WeyH49}, which allows to define a
zeta regularized determinant via
$${\det}_\zeta( \Delta) = e^{-\zeta ' (0,\Delta )}.$$
This definition has been used extensively in quantum field theory,
see
i.e.~\cite{buch92b,byts96-266-1,gerald,BElE-etc94,espo94b,espo97b,HawS77,BKirK01},
as well as in the context of the Reidemeister-Franz torsion
\cite{RS71,RaD-SiI73}. In particular, in one dimension rather
general and elegant results may be obtained, which has attracted
the interest of mathematicians especially in the last decade or so
\cite{BuD-FrL-KaT92,BFK2,dre78,for87,for92,Les,LesTol,lev77}. In
higher dimensions known results are restricted to highly symmetric
configurations
\cite{BKD,BGKE,bran94-344-479,byts96-266-1,dowk94-162-633,dowk94-11-557,dowk94-35-4989,dowk99-7-641}
or conformally related ones
\cite{blau88-310-163,blau89-4-1467,bran94-344-479,dowk95-12-1363,dowk96-5-799}.

Whereas most analysis has been done in the smooth setting,
relevant situations do not fall into this category. For example,
in order to compute quantum corrections to classical solutions in
Euclidean Yang-Mills theory \cite{call80-77-229,thoo76-14-3432}
singular potentials need to be considered. They also serve for the
description of physical systems like the Calogero Model
\cite{basu03-311-87,basu03-659-437,calo69-10-2191,calo69-10-2197,calo71-12-419,FPW,olsh81-71-313}
and conformal invariant quantum mechanical models
\cite{alfa76-34-569,BuW-GeF85,camb01-287-14,camb01-287-57,CoS-HoB02,FrW-LaD-SpR71,RadC75}.
More recently they became popular among physicists working on
space-times with horizons. There, for a variety of black holes,
singular potentials are used to describe the dynamics of quantum
particles in the asymptotic near-horizon region
\cite{birm01-505-191,clau98-81-4553,gibb99-454-187,govi00-583-291,more02-647-131}.

A similar situation occurs when manifolds are allowed to have
conical singularities \cite{Ch79,Ch}. Under these circumstances,
in general, $\zeta ' (0,\Delta )$ will not be defined, although
for special instances this definition still makes sense; nearly
all of the literature has concentrated on these special instances.
In order to describe these instances in more detail, let us
consider a bounded generalized cone. As we will see below, the
Laplacian on a bounded generalized cone has the form
$$\Delta = - \frac{\partial ^2}{\partial r^2} + \frac 1 {r^2}
A_\Gamma ,$$ where $A_\Gamma$ is defined on the base of the cone.
If $A_\Gamma$ has eigenvalues in the interval $[\frac 3 4,\infty
)$ only, one can show that $\Delta$ is essentially self-adjoint
and no choices for self-adjoint extensions exist. Spectral
functions, in particular the determinant, have been analyzed in
detail in \cite{BKD}. In case $A_\Gamma$ has one or more
eigenvalues in the interval $[-\frac 1 4,\frac 3 4)$ different
self-adjoint extensions exist; see for example \cite{MoE99}. Most
literature is concerned with the so-called Friedrich's extension
\cite{BS1,CaC83,CaC88,Ch,cogn94-49-1029,CoG-ZeS97,DowJ77,DowJ94,BLeM97,Les,SprM05}
and homogeneous or scale-invariant extensions
\cite{Ch,BLeM97,LMP}. Exceptions are \cite{FPW,FMPS,FMP} where
general self-adjoint extensions associated with one eigenvalue in
$[-\frac 1 4,\frac 3 4)$ have been considered. Only recently,
properties of spectral functions for arbitrary self-adjoint
extensions over the generalized cone have been understood
\cite{KLP}; a summary of the results is given in Section
\ref{sec2}. In particular, the zeta function is shown to have a
logarithmic branch point at $s=0$, in addition to the standard
simple pole at $s=0$. A natural construct for the determinant is
to subtract off these singular terms and to consider the
derivative of the finite remainder. This also is explained in
Section \ref{sec2}.

The details of the singular behavior as $s\to 0$, as well as of
the finite terms, strongly depend on the self-adjoint extension.
In Section \ref{sec3} we therefore briefly review the construction
of self-adjoint extensions on the generalized cone using the
Hermitian symplectic extension theory
\cite{HarM00,HarMI00,BHuV-PyJ80,KocA79,KocA90,KocA91,KoV-ScR99,INovS99,PavB87}.
This, finally, provides the set-up for the analysis of the zeta
function for arbitrary self-adjoint extensions. Even in the most
general case eigenvalues are determined by an implicit or
transcendental equation, a perfect starting point for the contour
integration method described in detail in
\cite{BKD1,BKD,BGKE,BKirK01,KM1,KM2}. This method allows us to
find the determinant for arbitrary self-adjoint extensions, the
main result, see Theorem \ref{t-det0}, being derived in Section
\ref{sec4}. In Section \ref{sec-special} we apply the answer for
the general case to certain natural self-adjoint extensions. The
conclusions provide a brief summary.

\section{Zeta functions on generalized cones and their
$\zeta$-determinants}\label{sec2}

In this section we review the notion of Laplace-type operators
over generalized cones and we discuss the pathological properties
of their zeta functions, which may have poles of arbitrary
multiplicity and countably many logarithmic singularities. We
state a natural procedure to define the $\zeta$-regularized
determinant and finally, we state the main formulas of this paper.

\subsection{Generalized cones and regular singular operators}

Let $\Gamma$ be a smooth $(n-1)$-dimensional compact manifold
(with or without boundary). Then the generalized cone with base
$\Gamma$, also called a cone over $\Gamma$, is the $n$-dimensional
manifold
\[
M = [0, R]_r \times \Gamma ,
\]
where $R > 0$ and the metric of $M$ is of the type $dr^2 + r^2 h$
with $h$ a metric over $\Gamma$. Let $E$ be a Hermitian vector
bundle over $M$ and let
\[ \Delta_M : C^\infty_c(M \setminus
\{0\}\times\Gamma , E) \to C^\infty_c(M \setminus  \{0\}\times
\Gamma , E)
\]
be a Laplace-type operator with the Dirichlet condition at $r = R$
having the form
\[
\Delta_M = - \partial_r^2 - \frac{n - 1}{r} \partial_r +
\frac{1}{r^2} \Delta_\Gamma,
\]
where $\Delta_\Gamma$ is a Laplace-type operator acting on
$C^\infty(\Gamma,E_\Gamma)$ where $E_\Gamma:=E|_\Gamma$; if
$\Gamma$ has a boundary we put Dirichlet conditions (for example)
at $\partial \Gamma$. By introducing a Liouville transformation,
we can write $\Delta_M$ in an equivalent way that is more
convenient for analysis. Writing $\phi \in L^2(M,E,r^{n-1} dr d
h)$ as
\begin{equation} \label{phiiso}
\phi = r^{- \frac{n-1}{2}} \widetilde{\phi},
\end{equation}
where $\widetilde{\phi}:= r^{\frac{n-1}{2}} \phi$, we have
\[
\int_{M} \langle \phi , \psi \rangle \, r^{n-1} d r d h = \int_{M}
\langle \widetilde{\phi} , \widetilde{\psi} \rangle \, d r d h,
\]
and a short computation shows that
\[
\Delta_M \phi = \left( - \partial_r^2 - \frac{n - 1}{r} \partial_r
+ \frac{1}{r^2} \Delta_\Gamma \right) \phi = r^{- \frac{n-1}{2}}
\Delta\, \widetilde{\phi},
\]
where
\begin{equation} \label{Delta}
\Delta : = - \partial_r^2 + \frac{1}{r^{2}} A_{\Gamma}
\end{equation}
with $A_\Gamma :=  \Delta_\Gamma + \frac{n-1}{2}
\left(\frac{n-1}{2} - 1\right)$. In conclusion: Under the
isomorphism \eqref{phiiso}, $L^2(M,E,r^{n-1} dr dh)$ is identified
with $L^2(M,E)$ with the standard measure $dr dh$, and $\Delta_M$
is identified with the operator $\Delta$ in \eqref{Delta}. It
turns out that for analytical purposes, the operator $\Delta$ is
somewhat more natural to work with. Notice that if $\Delta_\Gamma$
happens to be nonnegative, then
\[
A_\Gamma = \Delta_\Gamma + \frac{n-1}{2} \left(\frac{n-1}{2} -
1\right) \geq \Delta_\Gamma - \frac14 \geq - \frac14,
\]
where we used the fact that the function $x ( x - 1)$ has the
minimum value $- \frac14$ (when $x = \frac12$). In fact, it is
both a necessary and sufficient condition that $A_\Gamma \geq -
\frac14$ in order that $\Delta_M$ (or $\Delta$) be bounded below
\cite{BS1,CaC83,CaC88}. For this reason, we henceforth
\emph{assume} that $A_\Gamma \geq - \frac14$. The operator
$\Delta$ is called a second order \emph{regular singular operator}
\cite{BS1}.

Let $\{\la_\ell\}$ denote the spectrum of $A_\Gamma$.  Then Weyl's
alternative \cite{weyl10-68-220} immediately shows that $0$ is in
the limit case if and only if $-1/4 \leq \lambda_\ell < 3/4$
\cite{weidb87}. Consider only those eigenvalues in $[- \frac14,
\frac34)$:
\begin{equation} \label{laq}
-\frac14 = \underbrace{\lambda_1 = \lambda_2 = \cdots =
\lambda_{q_0}}_{ = - \frac14} < \underbrace{\lambda_{q_0+1} \leq
\lambda_{q_0 + 2} \leq \cdots \leq \la_{q_0 + q_1}}_{- \frac14 <
\la_\ell < \frac34} ,
\end{equation}
where each eigenvalue is counted according to its multiplicity.
Then, as a consequence of von Neumann's theory of self-adjoint
extensions the self-adjoint extensions of $\Delta$ are in a
one-to-one correspondence to the Lagrangian subspaces in $\C^{2q}$
where $q = q_0 + q_1$ and where $\C^{2q}$ has the symplectic form
described in \eqref{HSS}
\cite{ChJ79,Ch1,Ch,GM,BLeM97,MoE99,KoV-ScR99}. A concrete
description of these Lagrangian subspaces is as follows (see
Proposition \ref{p:relation}). A subspace $L \subset \C^{2q}$ is
Lagrangian if and only if there exists $q \times q$ complex
matrices $\Aa$ and $\Bb$ such that the rank of the $q \times 2 q$
matrix $\begin{pmatrix} \Aa & \Bb
\end{pmatrix}$ is $q$, $\Aa' \, \Bb^*$ is self-adjoint where
$\Aa'$ is the matrix $\Aa$ with the first $q_0$ columns multiplied
by $-1$, and
\begin{equation} \label{LABLagrang}
L = \{ v \in \C^{2q} \, | \, \begin{pmatrix} \Aa & \Bb
\end{pmatrix} v = 0\}.
\end{equation}
Given such a subspace $L \subset \C^{2q}$ there exists a
canonically associated domain $\mathfrak{D}_L \subset H^2(M,E)$
such that
\[
\Delta_L := \Delta : \mathfrak{D}_L \longrightarrow L^2(M,E)
\]
is self-adjoint (see Proposition \ref{p:sae}).

\subsection{Exotic zeta functions}
$\Delta_L$ has pure discrete spectrum \cite{BLeM97}, and hence, if
$\{\mu_j\}$ denotes the spectrum of $\Delta_L$, then we can form
the zeta function
\[
\zeta(s, \Delta_L) := \sum_{\mu_j \ne 0} \frac{1}{\mu_j^s} .
\]
The meromorphic structure of $\zeta(s, \Delta_L)$ (or the
corresponding heat trace) has been extensively studied for special
self-adjoint extensions, as for example the Friedrichs extension
\cite{BKD,BS1,CaC83,CaC88, Ch,cogn94-49-1029,CoG-ZeS97,DowJ77,
DowJ94,FurD94,SprM05}, which corresponds to taking $\Aa = 0$ and
$\Bb = \Id$ in \eqref{LABLagrang} \cite{BS1}, and the homogeneous
or scale-invariant extensions \cite{Ch,BLeM97,LMP}, which
corresponds to taking $\Aa$ and $\Bb$ to be diagonal matrices with
$0$'s and $1$'s along the diagonal such that the first $q_0$
entries along the diagonal of $\Bb$ are $1$'s and $\Aa + \Bb =
\Id$ \cite{LMP}. In these cases, the zeta function has the
``regular" meromorphic structure; that is, the same structure as
on a smooth manifold with one exception, $\zeta(s,\Delta_L)$ might
have a pole at $s = 0$. For general self-adjoint extensions, the
meromorphic structure has been studied in
\cite{FPW,FMPS,FMP,KLP,MoE99}. The papers \cite{FPW,FMPS,FMP} are
devoted to one-dimensional Laplace-type operators over the unit
interval and \cite{KLP,MoE99} study the general case of operators
over manifolds. The papers \cite{FPW,FMPS,FMP,MoE99} show that
$\zeta(s,\Delta_L)$ has, in addition to the ``regular" poles,
additional simple poles at ``unusual" location. In \cite{KLP} it
was shown that the zeta function $\zeta(s, \Delta_L)$ has, in the
general case, in addition to the ``unusual" poles, meromorphic
structures that remained unobserved and which are unparalleled in
the zeta function literature such as poles of arbitrary order and
logarithmic singularities.

The main result of \cite{KLP} not only states the existence of
such exotic singularities but it also gives an
algebraic-combinatorial algorithm that finds these singularities
explicitly. Although the algorithm is described in detail there,
we have to provide a summary in order to set up the notation used
in the rest of the paper. The algorithm is described as follows.

{\bf Step 1:} Let $\Aa$ and $\Bb$ be as in \eqref{LABLagrang} and
define the function
\begin{equation} \label{polyp}
p(x,y) := \det
\begin{pmatrix} \Aa & \Bb\\
\begin{array}{cccc}
x\, \Id_{q_0} & 0 & 0 & 0\\
0 & \tau_1\, y^{2 \nu_1} & 0 & 0\\
0 & 0 & \ddots & 0\\
0 & 0 & 0 & \tau_{q_1}\, y^{2 \nu_{q_1}}
\end{array}
&  \Id_q &
\end{pmatrix},
\end{equation}
where $\Id_k$ denotes the $k \times k$ identity matrix and where
\[
\nu_j := \sqrt{\la_{q_0 + j} + \frac14}\ , \quad \tau_j =
2^{2\nu_j} \frac{ \Gamma(1 + \nu_j)}{ \Gamma(1 - \nu_j)},\qquad j
= 1,\ldots, q_1 ,
\]
with $q_0,q_1,\la_j$ as in \eqref{laq}. Expanding the determinant,
we can write $p(x,y)$ as a finite sum
\[
p(x,y) = \sum a_{j \alpha} \, x^j\, y^{2 \alpha} ,
\]
where the $\alpha$'s are linear combinations of $\nu_1, \ldots,
\nu_{q_1}$ and the $a_{j \alpha}$'s are constants. Let $\alpha_0$
be the smallest of all $\alpha$'s with $a_{j \alpha} \ne 0$ and
let $j_0$ be the smallest of all $j$'s amongst the $a_{j \alpha_0}
\ne 0$. Then factoring out the term $a_{j_0 \alpha_0}\, x^{j_0}\,
y^{2 \alpha_0}$ in $p(x,y)$ we can write $p(x,y)$ in the form
\begin{equation} \label{pxy}
p(x,y) = a_{j_0 \alpha_0}\, x^{j_0}\, y^{2 \alpha_0} \Big( 1 +
\sum  b_{k \beta} \, x^k \, y^{2 \beta} \Big)
\end{equation}
for some constants $b_{k \beta}$ (equal to $a_{k \beta}/a_{j_0
\alpha_0}$).

{\bf Step 2:} Second, putting $z = \sum b_{k \beta} \, x^k \, y^{2
\beta}$ into the power series $\log (1 + z) = \sum_{k = 1}^\infty
\frac{(-1)^{k-1}}{k} z^k$ and {\it formally} expanding, we can
write
\begin{equation} \label{clx}
\log \Big( 1 + \sum  b_{k \beta} x^k y^{2 \beta} \Big) = \sum
c_{\ell \xi} \, x^\ell \, y^{2 \xi}
\end{equation}
for some constants $c_{\ell \xi}$. By construction, the $\xi$'s
appearing in \eqref{clx} are nonnegative, countable, and approach
$+\infty$ unless $\beta = 0$ is the only $\beta$ in \eqref{pxy},
in which case only $\xi = 0$ occurs in \eqref{clx}. Also, for a
fixed $\xi$, the $\ell$'s with $c_{\ell \xi} \ne 0$ are bounded
below.

{\bf Step 3:} Third, for each $\xi$ appearing in \eqref{clx},
define
\begin{equation} \label{ellxi}
p_\xi := \min \{\ell \leq 0 \, |\, c_{\ell \xi} \ne 0\}\qquad
\text{and}\qquad \ell_\xi := \min \{\ell > 0 \, |\, c_{\ell \xi}
\ne 0\},
\end{equation}
whenever the sets $\{\ell \leq 0 \, |\, c_{\ell \xi} \ne 0\}$ and
$\{\ell > 0 \, |\, c_{\ell \xi} \ne 0\}$, respectively, are
nonempty. Let ${\mathscr P}$, respectively ${\mathscr L}$, denote
the set of $\xi$ values for which the respective sets are
nonempty. The following theorem is our main result \cite[Th.\
2.1]{KLP}.

\begin{theorem}
\label{thm-main} The $\zeta$-function $\zeta(s,\Delta_L)$ extends
from $\Re s > \frac{n}{2}$ to a meromorphic function on $\C
\setminus (-\infty,0]$. Moreover, $\zeta(s,\Delta_L)$ can be
written in the form
\[
\zeta(s,\Delta_L) = \zeta_{\mathrm{reg}}(s,\Delta_L) +
\zeta_{\mathrm{sing}}(s,\Delta_L),
\]
where $\zeta_{\mathrm{reg}}(s,\Delta_L)$ has possible ``regular"
poles at the ``usual" locations $s = \frac{n - k}{2}$ with $s
\notin -\N_0$ for $k \in \N_0$ and at $s = 0$ if $\dim \Gamma >
0$, and where $\zeta_{\mathrm{sing}}(s,\Delta_L)$ has the
following expansion:
\begin{multline} \label{zetasing}
\zeta_{\mathrm{sing}}(s,\Delta_L) = \frac{\sin (\pi s)}{\pi}
\bigg\{ (j_0 - q_0) e^{-2 s (\log 2 - \gamma)} \log s\\ + \sum_{
\xi \in {\mathscr P}} \frac{f_\xi(s)}{(s + \xi)^{|p_\xi| + 1}} +
\sum_{ \xi \in {\mathscr L}} g_\xi(s) \log (s + \xi) \bigg\},
\end{multline}
where $j_0$ appears in \eqref{pxy} and $f_\xi(s)$ and $g_\xi(s)$
are entire functions of $s$ such that
\[
f_\xi(-\xi) = (-1)^{|p_\xi|+1} c_{p_\xi \xi} \,
\frac{|p_\xi|!}{2^{|p_\xi|}} \, \xi
\]
and
\[
g_\xi(s) = \begin{cases} c_{\ell_0, 0} \, \frac{2^{\ell_0}}{
(\ell_0 - 1)!} s^{\ell_0} + \mathcal{O}(s^{\ell_0 + 1}) & \text{if
$\xi = 0$,}\\[.1cm] - c_{\ell_\xi \xi} \, \frac{\xi\,\,
2^{\ell_\xi}}{(\ell_\xi - 1)!} (s + \xi)^{\ell_\xi - 1} +
\mathcal{O}((s + \xi)^{\ell_\xi}) & \text{if $\xi > 0$.}
\end{cases}
\]
\end{theorem}

\begin{remark} The expansion \eqref{zetasing} means that for any $N \in \N$,
\begin{multline*}
\zeta_{\mathrm{sing}}(s,\Delta_L) = \frac{\sin (\pi s)}{\pi}
\bigg\{ (j_0 - q_0) e^{-2 s (\log 2 - \gamma)} \log s  + \sum_{
\xi \in {\mathscr P} ,\, \xi \leq N} \frac{f_\xi(s)}{(s +
\xi)^{|p_\xi| + 1}} \\ + \sum_{ \xi\in {\mathscr L} ,\, \xi \leq
N} g_\xi(s) \log (s + \xi) \bigg\} + F_N(s),
\end{multline*}
where $F_N(s)$ is holomorphic for $\Re s \geq - N$. Note that the
leading terms as $s\to 0$ are contained in
$\zeta_{\mathrm{reg}}(s,\Delta_L)$ and the first term of
$\zeta_{\mathrm{sing}}(s,\Delta_L)$.
\end{remark}

\subsection{$\zeta$-determinant formul\ae}

For a general self-adjoint extension, Theorem \ref{thm-main} shows
that the $\zeta(s,\Delta_L)$ may not only have a simple pole at $s
= 0$ (from $\zeta_{\mathrm{reg}}(s,\Delta_L)$) but also a
logarithmic singularity at $s = 0$. Needless to say, the zeta
function is rarely regular at $s = 0$ except for special
self-adjoint extensions. In particular, the usual definition of
the zeta-regularized determinant is ill-defined via taking the
derivative of $\zeta(s,\Delta_L)$ at $s = 0$. However, we can
still associate a natural definition of a determinant by
subtracting off the singularities. Thus, let us define
\[
\zeta_0(s, \Delta_L) := \zeta(s, \Delta_L) \ -\ \frac{c}{s} \ - \
(j_0 - q_0) s \log s ,
\]
where $c = \mathrm{Res}_{s = 0} \zeta_{\mathrm{reg}}(s,\Delta_L)$.
The term $c/s$ cancels the possible pole of
$\zeta_{\mathrm{reg}}(s,\Delta_L)$ at $s = 0$ and by the explicit
formula \eqref{zetasing} for $\zeta_{\mathrm{sing}}(s,\Delta_L)$,
the term $(j_0 - q_0) s \log s$ cancels the logarithmic
singularity of $\zeta_{\mathrm{sing}}(s,\Delta_L)$ at $s = 0$ up
to a term that is $\mathcal{O}(s^2 \log s)$ at $s = 0$. It follows
that $\lim_{s \to 0^+} \zeta_0'(s, \Delta_L)$ exists. Therefore,
we can define
\[
{\det}_\zeta (\Delta_L) :=  \exp \left( - {\displaystyle \lim_{s
\to 0^+}} \zeta_0'(s, \Delta_L) \right)\label{defdet}
\]
This definition of course agrees with the standard definition in
case $\zeta(s, \Delta_L)$ is regular at $s = 0$. In Theorem
\ref{t-det0} below, we find an explicit formula for this
determinant. Because of some unyielding constants, it is elegant
to write our main formula as a relative formula in terms of the
Neumann extension. The \emph{Neumann extension} is given by
choosing $\Aa$ and $\Bb$ to be the diagonal matrices with the $q_0
+ 1, \ldots, q$ entries in $\Aa$ equal to $1$ and the
$1,\ldots,q_0$ entries in $\Bb$ equal to $1$ with the rest of the
entries $0$. By Corollary \ref{c:Neumann} (or \cite{LMP}), we find
the explicit formula
\begin{equation} \label{detDN}
{\det}_\zeta (\Delta_\Nn ) = (2 \pi R)^{\frac{q}{2}} \prod_{j =
1}^{q_1} \frac {2^{\nu _{j}} \, R^{-\nu_j}}{\Gamma (1 - \nu _{j}
)} \cdot {\det}_\zeta(\widetilde{\Delta})
\end{equation}
where $\widetilde{\Delta}$ is the (essentially self-adjoint)
operator obtained by projecting $\Delta$ onto the eigenvalues of
$A_\Gamma$ in $[\frac34,\infty)$ (see \eqref{def-tildeDel} for a
more precise definition of $\widetilde{\Delta}$). The determinant
${\det}_\zeta(\widetilde{\Delta})$ is given explicitly in Equation
(9.8) of \cite{BKD} when $R = 1$, with a similar formula holding
for arbitrary $R > 0$. We refer the reader to \cite{BKD} for the
appropriate details on ${\det}_\zeta(\widetilde{\Delta})$. The
following theorem is our main result.

\begin{theorem}\label{t-det0}
For a Lagrangian $L \subset \C^{2q}$ such that the operator
obtained by projecting $\Delta$ onto the eigenvalues of $A_\Gamma$
in $[-\frac14,\frac34)$ is invertible, we have
\[
\frac{{\det}_\zeta (\Delta_L )}{{\det}_\zeta(\Delta_\Nn)} =
\frac{(- 2 e^{\gamma})^{q_0 - j_0}}{a_{j_0 \alpha_0}} \, {\det
\begin{pmatrix} \Aa & \Bb \\
\begin{array}{cc} \Id_{q_0} & 0 \\
0 & {\mathbf R}^{2\nu} \end{array} &
\begin{array}{cc} (\log R) \Id_{q_0} & 0 \\
0 & \Id_{q_1} \end{array}
\end{pmatrix}
} ,
\]
where $a_{j_0 \alpha_0}$ is the coefficient in \eqref{pxy} and
$\mathbf{R}^{2\nu}$ is the $q_1 \times q_1$ diagonal matrix with
entries $R^{2 \nu_\ell}$ for $1\leq \ell \leq q_1$.
\end{theorem}

Combining this formula with \eqref{detDN}, we get an explicit
formula for ${\det}_\zeta (\Delta_L )$.

%
The next result follows from an application of
Theorem \ref{t-det0} to a particular class of matrices $\Aa$ and
$\Bb$.

\begin{theorem}\label{t-det}
Let $q - r = \mathrm{rank}(\Aa)$ and assume that $\Aa$ has $r$
rows and columns identically zero. Let $i_1,..., i_q$ be a
permutation of the numbers $1,..., q$ such that the rows and
columns $i_1,...,i_{r}$ of $\Aa$ are zero. Choose $j_0 \in
\{0,1,\ldots, q_0\}$ such that
\[
1 \leq i_1 < i_2 < \cdots < i_{j_0} \leq q_0 < i_{j_0 + 1} <
\cdots < i_r \leq q .
\]
Let $I_{r}$ denote the $q \times q$ matrix which is zero
everywhere except along the diagonal where the entries
$i_1,...,i_{r}$ equal $1$, and let $I_{q-r}$ denotes the $q \times
q$ matrix which is zero everywhere except along the diagonal where
the entries $i_{r+1},...,i_q$ equal $1$. Then for a Lagrangian $L$
having $\Aa$ as a first component and satisfying the condition in
Theorem \ref{t-det0}, we have:
\begin{multline*}
\frac{{\det}_\zeta (\Delta_L )}{{\det}_\zeta(\Delta_\Nn)} = (- 2
e^{\gamma})^{q_0 - j_0} \prod_{j=j_0 + 1}^{r} \left[ 2^{-2 \nu
_{i_j}} \frac {\Gamma (1 - \nu _{i_j} )}{\Gamma (1+\nu _{i_j} )}
\right] {\det \left(
\begin{array}{cc}
\Aa & \Bb \\
I_{r} & I_{q - r} \end{array} \right)}^{-1} \times
\\
{\det
\begin{pmatrix} \Aa & \Bb \\
\begin{array}{cc} \Id_{q_0} & 0 \\
0 & {\mathbf R}^{2 \nu} \end{array} &
\begin{array}{cc} (\log R) \Id_{q_0} & 0 \\
0 & \Id_{q_1} \end{array}
\end{pmatrix}
}   .
\end{multline*}
\end{theorem}

See Section \ref{sec-special} for more special cases including
one-dimensional operators.

\section{The Hermitian symplectic theory of self-adjoint
extensions}\label{sec3}
In this section we briefly explain the correspondence between
self-adjoint extensions and the Lagrangian subspaces described by
(\ref{LABLagrang}). This correspondence is a direct consequence of
von Neumann's classical theory of self-adjoint extensions; a
partial list of relevant references is
\cite{ChJ79,Ch1,Ch,GM,BLeM97,MoE99,KoV-ScR99,HarM00,HarMI00,BHuV-PyJ80,KocA79,KocA90,KocA91,KoV-ScR99,INovS99,PavB87,weidb80,weidb87}.
\subsection{Reduction to the model problem}
\label{sec-modelproblem}

Let $\{\lambda_\ell\}$ denote the set of all eigenvalues of
$A_\Gamma$ and let $E_\ell$ denote the span of the $\la_\ell$-th
eigenvector. Let $\Pi$ and $\Pi^\perp$ denote, respectively, the
orthogonal projections of $L^2(\Gamma,E_\Gamma)$ onto $W : =
\bigoplus_{-\frac14 \leq \lambda_\ell < \frac34} E_\ell \cong
\C^q$ and $W^\perp$. Using the isometry between
\[
L^2([0,R]\times \Gamma, E) \cong L^2([0,R], L^2(\Gamma,
E_\Gamma)),
\]
we obtain the corresponding projections on $L^2([0,R]\times
\Gamma, E)$, which we denote with the same notations $\Pi$ and
$\Pi^\perp$. Since $A_\Gamma$ preserves $W$ and $W^\perp$, we can
write
\[
\Delta =  \Ll \oplus \widetilde{\Delta} ,
\]
where
\begin{equation}\label{def-tildeDel}
\widetilde{\Delta} := \Pi^\perp \Delta \Pi^\perp = - \partial_r^2
+ \frac{1}{r^{2}} A_\Gamma \big|_{W^\perp} ,
\end{equation}
and $\Ll$ is the (matrix) ordinary differential operator
\[
\Ll := \Pi \Delta \Pi = - \frac{d^2}{dr^2} + \frac{1}{r^2} A ,
\]
where $A$ is the $q \times q$ diagonal matrix
\[
A = \begin{pmatrix} - \frac14 \Id_{q_0}& 0 \\
0 & \begin{array}{ccccc}
\la_{q_0 + 1} & 0 & 0 & \cdots & 0\\
0 & \la_{q_0 + 2} & 0 & \cdots & 0\\
0 & 0 & \la_{q_0 + 3} & \cdots & 0 \\
0 & 0 & 0 & \ddots & 0\\
0 & 0 & 0 & \cdots & \la_{q_0 + q_1}
\end{array}
\end{pmatrix} ;
\]
here we write $A$ with respect to the basis of $W =
\bigoplus_{-\frac14 \leq \lambda_\ell < \frac34} E_\ell \cong
\C^q$. It is well-known that the operator $\widetilde{\Delta}$ is
essentially self-adjoint \cite{BS1,BS2,CaC83,CaC88,MoE99}
Therefore, the various self-adjoint extensions of
$\Delta$ are simply the various self-adjoint extensions of the
``toy model operator" $\Ll$, which we now study.

\subsection{Self-adjoint extensions of the model
operator}\label{subs:selfadjoint}
The key to determining the self-adjoint extensions of $\Ll$ is to
first characterize the \emph{maximal domain} of $\Ll$:
\[
\mathfrak{D}_\max := \big\{\, \phi \in L^2([0,R], \C^q) \, |\, \Ll
\phi \in L^2([0,R], \C^q) \ \text{and}\ \phi(R) = 0\, \big\},
\]
which is the largest set of $L^2$ functions on which $\Ll$ can act
and stay within $L^2$. As an immediate consequence of Cheeger
\cite{Ch1,Ch} we have
\begin{proposition} \label{p:maxdommodel}
$\phi \in \mathfrak{D}_\max$ if and only if $\phi(R)=0$ and $\phi$
has the following form:
\begin{align} \label{phiasymp1model}
\phi =& \sum_{\ell = 1}^{q_0} \left\{ {c_\ell(\phi)} \, r^{
\frac12} e_\ell + {c_{q + \ell}(\phi)} \,
r^{ \frac12}\log r e_\ell \right\}\\
&\qquad + \sum_{\ell = 1}^{q_1} \left\{ c_{q_0 + \ell}(\phi) \,
r^{\nu_\ell + \frac12} e_{q_0 + \ell} + c_{q + q_0 + \ell}(\phi)
\, r^{-\nu_\ell + \frac12} e_{q_0 + \ell} \right\} +
\widetilde{\phi} ,\notag
\end{align}
where
\[
\nu_\ell := \sqrt{\la_{q_0 + \ell} + \frac14}\ >\ 0,
\]
$e_\ell$ is the column vector with $1$ in the $\ell$-th slot and
$0$'s elsewhere, the $c_j(\phi)$'s are constants, and the
$\widetilde{\phi}$ is continuously differentiable on $[0,R]$ such
that $\widetilde{\phi}(r)=\mathcal{O}(r^{\frac32})$ and
$\widetilde{\phi}'(r) = \mathcal{O}(r^{\frac12})$ near $r=0$, and
$\Ll \widetilde{\phi} \in L^2([0,R],\C^q)$.
\end{proposition}
We next want to formulate the correspondence between self-adjoint
extensions and Lagrangian subspaces with respect to a suitable
symplectic form. Let
\[
J : = \begin{pmatrix} 0 & -\Id_q \\  \Id_q & 0 \end{pmatrix},
\]
and recall that
\[
\C^{2q} \times \C^{2q} \ni (v,w) \mapsto \langle J v , w \rangle
\in \C
\]
is the standard Hermitian symplectic form on $\C^{2q}$; that is,
this form is Hermitian antisymmetric and nondegenerate. Now
defining $T : \C^{2q} \to \C^{2q}$ by
\[
T(v_1,\ldots, v_{2q}) = (-v_1,\ldots, - v_{q_0},v_{q_0 + 1} ,
\ldots , v_{2q})
\]
and putting $\vec{\phi} = (c_1(\phi), c_2(\phi), \ldots,
c_{2q}(\phi))^t$, $\vec{\psi}= (c_1(\psi), c_2(\psi), \ldots,
c_{2q}(\psi))^t$, one has
\begin{equation} \label{HSS}
\langle \Ll \phi , \psi \rangle - \langle \phi , \Ll \psi \rangle
= \langle J T \vec{\phi} , T \vec{\psi} \, \rangle =:
\omega(\vec{\phi}, \vec{\psi}),
\end{equation}
where
\[
\omega(v,w) := \langle J T v ,T w \rangle \quad \text{for all\ }\
v,w \in \C^{2q}
\]
defines a symplectic form on $\C^{2q}$. We say that a subspace $L
\subset \C^{2q}$ is \emph{Lagrangian} (with respect to $\omega$)
if
\[
\{ w \in \C^{2q} \ |\ \omega(v,w) = 0 \ \ \text{for all}\ \ v \in
L \} = L.
\]
Self-adjoint extensions of $\Ll$ are then in one-to-one
correspondence with Lagrangian subspaces of $(\C^{2q},\omega)$ in
the sense that given any Lagrangian subspace $L \subset \C^{2q}$
and defining
\[
\mathfrak{D}_L := \{ \phi \in \mathfrak{D}_\max \ |\ \vec{\phi}
\in L\},
\]
the operator
\[
\Ll_L := \Ll : \mathfrak{D}_L \longrightarrow L^2([0,R], \C^{q})
\]
is self-adjoint and any self-adjoint extension of $\Ll$ is of the
form $\Ll_L$ for some Lagrangian subspace $L \subset \C^{2q}$.
The fact that any Lagrangian subspace $L\subset \C^{2q}$ with
respect to the standard symplectic form can be described by a
system of equations
\begin{equation} \label{Lag}
L = \left\{ v \in \C^{2 q} \, |\, \begin{pmatrix} \Aa & \Bb
\end{pmatrix} v = 0 \right\} \subset \C^{2q},
\end{equation}
where $\Aa$ and $\Bb$ are $q \times q$ matrices such that
$\begin{pmatrix} \Aa & \Bb \end{pmatrix}$ has full rank and $\Aa
\, \Bb^*$ is self-adjoint translates into the following result
when the symplectic form $\omega$ is used.
\begin{proposition}\label{p:relation} The set in \eqref{Lag} is a
Lagrangian subspace of $(\C^{2q}, \omega)$ if and only if the rank
of $\begin{pmatrix} \Aa & \Bb \end{pmatrix}$ is $q$ and $\Aa' \,
\Bb^*$ is self-adjoint where $\Aa'$ is the matrix $\Aa$ with the
first $q_0$ columns of $\Aa$ multiplied by $-1$.
\end{proposition}

The following proposition concludes our summary of basically known
results.

\begin{proposition} \label{p:sae}
The self-adjoint extensions of $\Delta$ are in one-to-one
correspondence with Lagrangian subspaces of $(\C^{2q},\omega)$.
More, precisely, self-adjoint extensions are of the form
\[
\Delta_L = \Ll_L \oplus \widetilde{\Delta},
\]
where
\[
\Ll_L := \Ll : \mathfrak{D}_L \to L^2([0,R], \C^{q})\ , \quad
\mathfrak{D}_L := \{ \phi \in \mathfrak{D}_\max \ |\ \vec{\phi}
\in L\} .
\]
Here, $L \subset \C^{2q}$ is given by \eqref{Lag} where $\Aa$ and
$\Bb$ are $q \times q$ matrices such that $\begin{pmatrix} \Aa &
\Bb \end{pmatrix}$ has rank $q$ and $\Aa' \, \Bb^*$ is
self-adjoint.
\end{proposition}

\section{Proof of Theorem \ref{t-det0}}\label{sec4}

In this section we prove Theorem \ref{t-det0} using the contour
integration method \cite{BKD1,BKD,BGKE,BKirK01,KM1,KM2}. We begin
by reducing our computation to the model operator.

\subsection{Reduction to the model problem}
\label{sec-modelproblem2}
From the results in Section \ref{sec3} it is clear that the zeta
function of $\Delta_L$ splits according to
\begin{equation} \label{zetaDL}
\zeta(s, \Delta_L) = \zeta_{\mathrm{reg}}(s,\Delta_L) +
\zeta_{\mathrm{sing}}(s,\Delta_L),
\end{equation}
where
\[
\zeta_{\mathrm{reg}}(s,\Delta_L) := \zeta (s,
\widetilde{\Delta})\quad \text{and}\quad
\zeta_{\mathrm{sing}}(s,\Delta_L) := \zeta(s, \Ll_L) .
\]
The properties of $\widetilde{\Delta}$, including the spectral
functions, have been studied extensively, see for example
\cite{BKD,Ch,CoG-ZeS97,DowJ94}. In particular,
$\zeta_{\mathrm{reg}}(s,\Delta_L)$ has possible poles at the usual
locations $s = \frac{n - k}{2}$ with $s \notin -\N_0$ for $k \in
\N_0$ and at $s = 0$ if $\dim \Gamma > 0$.  The residue of
$\zeta_{\mathrm{reg}}(s,\Delta_L)$ at $s = 0$ is given by
\[
c := \mathrm{Res}_{s = 0} \zeta_{\mathrm{reg}}(s,\Delta_L) = -
\frac12 \mathrm{Res}_{s = - \frac12} \zeta (s, A_\Gamma).
\]
In particular, this vanishes if $\zeta (s, A_\Gamma )$ is in fact
analytic at $s = -\frac12$. Furthermore, the determinant
\[
{\det}_\zeta(\widetilde{\Delta}) := \exp \left( - {\displaystyle
\frac{d}{ds}} \bigg|_{s = 0} \bigg\{ \zeta(s, \widetilde{\Delta})
- \frac{c}{s} \bigg\} \right)
\]
is thoroughly studied in \cite{BKD}. The meromorphic structure of
the singular function $\zeta_{\mathrm{sing}}(s,\Delta_L) :=
\zeta(s, \Ll_L)$ has the properties stated in Theorem
\ref{thm-main}, which was proved in \cite{KLP}. In particular,
\[
\zeta_0(s, \Ll_L) := \zeta(s, \Ll_L) \ - \ (j_0 - q_0) s \log s ,
\]
is differentiable at $s=0$ and so
\[
{\det}_\zeta(\Ll_L) := \exp \left( - \lim_{s_0\to 0^+}
{\displaystyle \frac{d}{ds}}\bigg|_{s_0 = 0} \zeta_0(s, \Ll_L)
\right)
\]
is defined. Also, by \eqref{zetaDL}, we have
\[
{\det}_\zeta(\Delta_L) = {\det}_\zeta(\Ll_L) \cdot
{\det}_\zeta(\widetilde{\Delta})
\]
Therefore, we have reduced to computing ${\det}_\zeta(\Ll_L)$. We
shall compute this in Proposition \ref{prop-det}, but first we
need to review some fundamental results from \cite{KLP}.

\subsection{Properties of the implicit eigenvalue equation}

In order to analyze ${\det}_\zeta(\Ll_L)$, we need to understand
the behavior of the eigenvalue equation for $\Ll_L$. In order to
write down the eigenvalue equation, we need some notation. Define
the $q \times q$ matrices \tiny
\[
J_+(\mu) := \begin{pmatrix} J_0(\mu R) \Id_{q_0} & 0 & \cdots & 0 \\
0 & 2^{\nu_1} \Gamma(1 + \nu_1) \, \mu^{-\nu_1}
J_{\nu_1}(\mu R) & \cdots & 0 \\
0 &  0 & \cdots & 0\\
0 &  0 & \ddots & 0\\
0 &  0 & \cdots & 2^{\nu_{q_1}} \Gamma(1 + \nu_{q_1}) \,
\mu^{-\nu_{q_1}} J_{\nu_{q_1}}(\mu R) \end{pmatrix}
\]
\normalsize and \tiny
\[
J_-(\mu) := \begin{pmatrix} \widetilde{J}_0(\mu R) \Id_{q_0} & 0 & \cdots & 0\\
0 & 2^{-\nu_1} \Gamma(1 - \nu_1) \,
\mu^{\nu_1} J_{-\nu_1}(\mu R) & \cdots & 0 \\
0 & 0 & \cdots & 0\\
0 & 0 & \ddots & 0\\
0 & 0 & \cdots & 2^{-\nu_q} \Gamma(1 - \nu_q) \, \mu^{\nu_{q_1}}
J_{-\nu_{q_1}}(\mu R) \end{pmatrix}
\]
\normalsize where $J_v(z)$ denotes the Bessel function of the
first kind and
\begin{equation}\label{e:def}
\widetilde{J}_0(\mu r) := \frac{\pi}{2}  Y_0(\mu r) - (\log \mu -
\log 2 + \gamma)\, J_0(\mu r) ,
\end{equation}
with $Y_0(z)$ the Bessel function of the second kind. Now we
define
\begin{equation} \label{Fmu}
F(\mu) := \det \begin{pmatrix} \Aa & \Bb \\
J_+(\mu) & J_-(\mu) \end{pmatrix}.
\end{equation}
Then $F(\mu)$ is an even function of $\mu$. Indeed, to see this
observe that, by definition, $F(\mu)$ is expressed in terms of
$\mu^v J_{-v}(\mu R)$ with appropriate $v$'s and the function
$\widetilde{J}_0(\mu R)$. The following equation \cite[p.\
360]{BAbM-StI92}
\begin{equation}\label{e:Jasymp}
z^{-v} J_v(z) = \sum_{k = 0}^\infty \frac{(-1)^k z^{2k}}{2^{v + 2
k} k! \, \Gamma ( v + k + 1)}
\end{equation}
shows that $\mu^{-v} J_v(\mu R)$ is even while the equality
\cite[p.\ 360]{BAbM-StI92}:
\begin{equation}\label{e:near0}
\frac{\pi}{2} Y_0(z) = \big( \log z - \log 2 + \gamma \big) J_0(z)
- \sum_{k = 1}^\infty \frac{H_k (- \frac14 z^2 )^k}{(k!)^2} ,
\end{equation}
where $H_k := 1 + \frac12 + \cdots + \frac{1}{k}$, and the
definition of $\widetilde{J}_0(\mu r)$ in \eqref{e:def} show that
$\widetilde{J}_0(\mu R)$ is even.

The importance of $F(\mu)$ lies in the following Proposition.

\begin{proposition} \label{prop-Fmu}
$\mu^2$ is an eigenvalue of $\Ll_L$ if and only if $F(\mu) = 0$.
Moreover,
\[
F(0) = {\det
\begin{pmatrix} \Aa & \Bb \\
\begin{array}{cc} \Id_{q_0} & 0 \\
0 & {\mathbf R}^{\nu} \end{array} &
\begin{array}{cc} (\log R) \Id_{q_0} & 0 \\
0 & {\mathbf R}^{-\nu}  \end{array}
\end{pmatrix}
} ,
\]
where $\mathbf{R}^{\pm \nu}$ are the $q_1 \times q_1$ diagonal
matrices with entries $R^{\pm \nu_\ell}$ for $1\leq \ell \leq
q_1$.
\end{proposition}

The first statement is straightforward to prove by solving the
equation $(\Ll_L - \mu^2) \phi = 0$ for $\phi$ and using the fact
that $L = \left\{ v \in \C^{2 q} \, |\, \begin{pmatrix} \Aa & \Bb
\end{pmatrix} v = 0 \right\}$ and that $\phi \in \mathfrak{D}_L$. The
details are provided in Proposition 4.2 of \cite{KLP}. The formula
for $F(0)$ follows directly from Equations \eqref{e:def},
\eqref{e:Jasymp} and  \eqref{e:near0}.

The following lemma analyzes the asymptotics of $F(\mu)$ as $|\mu|
\to \infty$ and is proved in Proposition 4.3 of \cite{KLP}.

\begin{lemma} \label{lem-asympF}
Let $\Upsilon \subset \C$ be a sector (closed angle) in the
right-half plane. Then we can write
\begin{multline} \label{asymF2}
F(i x) = (2 \pi R)^{-\frac q2} \prod_{j = 1}^{q_1} 2^{-\nu _{j} }
\Gamma (1 - \nu _{j} ) \, x^{|\nu| -\frac q2}\, e^{q x R} \,
(\widetilde{\gamma} - \log x)^{q_0} \times \\ p\Big( \big(
\widetilde{\gamma} - \log x \big)^{-1}, x^{-1} \Big) \, \Big( 1 +
f(x) \Big) ,
\end{multline}
where $\widetilde{\gamma}=\log 2-\gamma$, $p(x,y)$ is the function
in \eqref{polyp},
 and where as $|x| \to \infty$
with $x \in \Upsilon$, $f(x)$ is a power series in $x^{-1}$ with
no constant term.
\end{lemma}

Using this lemma, we prove the following Proposition.

\begin{proposition}
\label{prop-asympF} Let $\Upsilon \subset \C$ be a sector in the
right-half plane. Then we can write
\begin{equation} \label{asymF}
F(i x) = C x^{|\nu| -\frac q2 - 2 \alpha_0}e^{q x R}
(\widetilde{\gamma} - \log x)^{q_0 - j_0} \, \Big( 1 + G(x) \Big)
,
\end{equation}
where
\begin{equation} \label{C}
C = a_{j_0 \alpha_0}(2 \pi R)^{-\frac q2} \prod_{j = 1}^{q_1}
2^{-\nu _{j} } \Gamma (1 - \nu _{j} ),
\end{equation}
with $a_{j_0 \alpha_0}$ the coefficient in \eqref{pxy}, and $G(x)
= \mathcal{O}\left( \frac{1}{\log x} \right)$ and $G'(x) =
\mathcal{O}\left( \frac{1}{x (\log x)^2}\right)$ as $|x| \to
\infty$ with $x \in \Upsilon$.
\end{proposition}
\begin{proof}
Recall that $\alpha_0$ is the smallest of all $\alpha$'s with
$a_{j \alpha} \ne 0$ and $j_0$ is the smallest of all $j$'s
amongst the $a_{j \alpha_0} \ne 0$ in the expression
\[
p(x,y) = \sum a_{j \alpha} \, x^j\, y^{2 \alpha} ,
\]
which is obtained by expanding the determinant in the definition
of $p(x,y)$. Factoring out $a_{j_0 \alpha_0}\, x^{j_0}\, y^{2
\alpha_0}$ in $p(x,y)$ we can write $p(x,y)$ in the form (see
\eqref{pxy})
\[
p(x,y) = a_{j_0 \alpha_0}\, x^{j_0}\, y^{2 \alpha_0} \Big( 1 +
\sum  b_{k \beta} \, x^k \, y^{2 \beta} \Big),
\]
where we may assume that all $b_{k \beta} \ne 0$. By definition of
$\alpha_0$, all the $\beta$'s in this expression are nonnegative
real numbers and the $k$'s can be nonpositive or nonnegative
integers except when $\beta = 0$, when the $k$'s can only be
positive by definition of $j_0$. Now observe that
\begin{equation} \label{pg}
p\Big( \big( \widetilde{\gamma} - \log x \big)^{-1}, x^{-1} \Big)
\\ = a_{j_0 \alpha_0}\, \big( \widetilde{\gamma} - \log x
\big)^{-j_0}\, x^{- 2 \alpha_0} \Big( 1 + g(x)\Big),
\end{equation}
where $g(x) = \sum  b_{k \beta} \, \big( \widetilde{\gamma} - \log
x \big)^{-k} \, x^{-2 \beta}$. Notice that as $x \to \infty$,
\[
\big( \widetilde{\gamma} - \log x \big)^{-k} = \mathcal{O}\left(
\frac{1}{\log x} \right) \ \ \text{for $k > 0$},
\]
and, because $\log x$ increases slower than any positive power of
$x$,
\[
\big( \widetilde{\gamma} - \log x \big)^{-k} \, x^{-2 \beta} =
\mathcal{O}\left( \frac{1}{\log x} \right) \ \ \text{for $k \in
\Z$ and $\beta > 0$.}
\]
Therefore, $g(x) = \mathcal{O}\left( \frac{1}{\log x} \right)$. A
similar argument shows that $g'(x) = \mathcal{O}\left( \frac{1}{x
(\log x)^2} \right)$. Finally, replacing the formula \eqref{pg}
into the formula \eqref{asymF2}, we obtain
\begin{align*}
F(i x) & \sim C x^{|\nu| -\frac q2 - 2 \alpha_0}e^{q x R}
(\widetilde{\gamma} - \log x)^{q_0 - j_0} \, \Big( 1 +
g(x) \Big) \Big( 1 + f(x) \Big)\\
& \sim C x^{|\nu| -\frac q2 - 2 \alpha_0}e^{q x R}
(\widetilde{\gamma} - \log x)^{q_0 - j_0} \, \Big( 1 + G(x) \Big)
,
\end{align*}
where $C$ is given in \eqref{C} and $G(x) = f(x) + g(x) + f(x)\,
g(x)$. The ``big-$\mathcal{O}$" properties of $g(x)$ we discussed
above and the fact that $f(x)$ is a power series in $x^{-1}$ with
no constant term shows that $G(x)$ has the desired properties.
\end{proof}

\subsection{Computation of ${\det}_\zeta(\Ll_L)$}

In order to facilitate the computation, we first need to establish
the following

\begin{lemma} \label{lem-logint} For any constants $c$ and
$|t|$ such that $\log |t| > c$,  we have
\[
\int_{|t|}^\infty x^{-2 s - 1} \frac{1}{c - \log x}\, d x = e^{- 2
s c} \log s + e^{- 2 s c} \Big( \gamma + \log (2 (\log |t| - c)) +
\mathcal{O}(s)\Big),
\]
where $\mathcal{O}(s)$ is an entire function of $s$ that is
$\mathcal{O}(s)$ at $s = 0$.
\end{lemma}
\begin{proof}
To analyze this integral we make the change of variables $u = \log
x - c$ or $x = e^c\, e^u$, and obtain
\[
\int_{|t|}^\infty x^{-2 s - 1} \frac{1}{c - \log x}\, d x = - e^{-
2 s c} \int_{\log |t| - c}^\infty e^{-2 s u} \frac{d u}{u} .
\]
Making the change of variables $y = 2 s u$, we get
\begin{align*}
\int_{|t|}^\infty x^{-2 s - 1} \frac{1}{c - \log x}\, d x & = -
e^{- 2 s c} \int_{2 s (\log |t| - c)}^\infty e^{- y} \frac{d
y}{y}\\ & = e^{- 2 s c} \Ei \big(-2 s (\log |t| - c)\big),
\end{align*}
where $\Ei(z) := - \int_{-z}^\infty e^{-y} \frac{d y}{y}$ is the
\emph{exponential integral} (see \cite[Ch.\ 5]{BAbM-StI92} or
\cite[Sec.\ 8.2]{BGrI-RyI00}). From \cite[p.\ 877]{BGrI-RyI00}, we
have
\[
\Ei(z) = \gamma + \log (- z) + \sum_{k = 1}^\infty \frac{z^k}{k
\cdot k!},
\]
therefore
\begin{align*}
\notag \int_{|t|}^\infty x^{-2 s - 1} \frac{1}{c - \log x}\, d x &
= e^{- 2 s c} \Big( \gamma + \log
(2 s (\log |t| - c)) + \mathcal{O}(s) \Big)\\
& = e^{- 2 s c} \log s + e^{- 2 s c} \Big( \gamma + \log (2 (\log
|t| - c)) + \mathcal{O}(s)\Big),
\end{align*}
where $\mathcal{O}(s)$ is an entire function of $s$ that is
$\mathcal{O}(s)$ at $s = 0$.
\end{proof}

We now compute ${\det}_\zeta (\Ll_L )$ explicitly.

\begin{proposition} \label{prop-det} If $\ker \Ll_L=\{0\}$,
\begin{multline*}
{\det}_\zeta (\Ll_L ) = \frac{(2 \pi R)^{\frac{q}{2}}}{a_{j_0
\alpha_0}} \prod_{j = 1}^{q_1} \frac {2^{\nu _{j}}}{\Gamma (1 -
\nu _{j} )} (- 2 e^{\gamma})^{q_0 - j_0} \times
\\
{\det
\begin{pmatrix} \Aa & \Bb \\
\begin{array}{cc} \Id_{q_0} & 0 \\
0 & {\mathbf R}^\nu \end{array} &
\begin{array}{cc} (\log R) \Id_{q_0} & 0 \\
0 & {\mathbf R}^{-\nu} \end{array}
\end{pmatrix}
} .
\end{multline*}
\end{proposition}
\begin{proof}
First, applying the Argument Principle (which is really a form of
Cauchy's formula) \cite[p.\ 123]{BCoJ78}, the $\z$-function of
$\Ll_L$ is given by
\[
\zeta(s , \Ll_L ) = \frac{1}{2 \pi i} \int_\gamma \mu^{-2 s}
\frac{d}{d \mu} \log F(\mu) d \mu = \frac{1}{2 \pi i} \int_\gamma
\mu^{-2 s} \frac{F'(\mu)}{F(\mu)} d \mu ,
\]
where $\gamma$ is a contour in the plane shown in Figure
\ref{fig-contourdet}.
\begin{figure}
\centering \includegraphics{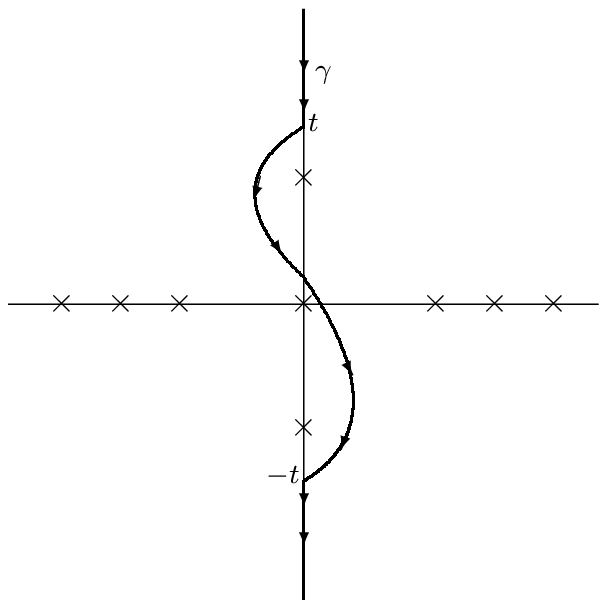} \caption{The
contour $\gamma$ for the zeta function. The $\times$'s represent
the zeros of $F(\mu)$ and squaring these $\times$'s are the
eigenvalues of $\Ll_L$. Here, $t$ is on the imaginary axis and
$|t|^2$ is larger than the largest absolute value of a negative
eigenvalue of $\Ll_L$ (if it has any). The contour $\gamma_t$ goes
from $t$ to $-t$.} \label{fig-contourdet}
\end{figure}
Breaking up our integral into three parts, one from $t$ to $i
\infty$, another from $-i \infty$ to $-t$, and then another over
$\gamma_t$, which is the part of $\gamma$ from $t$ to $-t$, we
obtain
\begin{align*}
\zeta(s,\Ll_L) & = \frac{1}{2 \pi i} \int_\gamma \mu^{-2 s}
\frac{d}{d\mu} \log F(\mu) \, d \mu\\
& = \frac{1}{2 \pi i} \bigg\{ - \int_{|t|}^\infty (i x)^{-2 s}
\frac{d}{d x} \log F(i x)\, d x + \int_{|t|}^\infty (-i x)^{-2 s}
\frac{d}{d x} \log F(-i x)\, d x \bigg\} \\
& \hspace{7.5cm} + \frac{1}{2 \pi i} \int_{\gamma_t} \mu^{-2 s}
\frac{F'(\mu)}{F(\mu)}\, d \mu\\ & = \frac{1}{2 \pi i} \Big( -
e^{- i \pi s} + e^{i \pi s}\Big) \int_{|t|}^\infty x^{-2 s}
\frac{d}{d x} \log F(i x)\, d x + \frac{1}{2 \pi i}
\int_{\gamma_t} \mu^{-2 s} \frac{F'(\mu)}{F(\mu)}\, d \mu ,
\end{align*}
or,
\begin{equation}\label{zetafn1}
\zeta(s,\Ll_L) = \frac{\sin \pi s}{\pi} \int_{|t|}^\infty x^{-2 s}
\frac{d}{d x} \log F(i x)\, d x + \frac{1}{2 \pi i}
\int_{\gamma_t} \mu^{-2 s} \frac{F'(\mu)}{F(\mu)}\, d \mu .
\end{equation}
The first step to compute $\zd (\Ll_L)$ is to construct the
analytical continuation of the first integral in \eqref{zetafn1}
to $s = 0$; the second term (being entire since it is an integral
over a finite contour) is already regular at $s = 0$. To do so,
recall Proposition \ref{prop-asympF} (see \eqref{asymF}), which
states that we can write
\[
F(i x) = C x^{|\nu| -\frac q2 - 2 \alpha_0}e^{q x R}
(\widetilde{\gamma} - \log x)^{q_0 - j_0} \, \Big( 1 + G(x) \Big)
,
\]
where
\[
C = a_{j_0 \alpha_0}(2 \pi R)^{-\frac q2} \prod_{j = 1}^{q_1}
2^{-\nu _{j} } \Gamma (1 - \nu _{j} ),
\]
and where $G(x) = \mathcal{O}\left( \frac{1}{\log x} \right)$  and
$G'(x) = \mathcal{O}\left( \frac{1}{x (\log x)^2}\right)$ as $|x|
\to \infty$. Hence,
\begin{align*}
\int_{|t|}^\infty x^{-2 s} \frac{d}{d x} \log F(i x)\, d x & =
\int_{|t|}^\infty x^{-2 s} \frac{d}{d x} \log \Big( 1 + G(x)
\Big)\, d x \\ & + \int_{|t|}^\infty x^{-2 s} \frac{d}{d x} \log
\Big( C x^{|\nu| -\frac q2 - 2 \alpha_0}e^{q x R}
(\widetilde{\gamma} - \log x)^{q_0 - j_0}  \Big)\, d x .
\end{align*}
The second integral can be computed explicitly:
\begin{align*}
\int_{|t|}^\infty & x^{-2 s} \frac{d}{d x} \log \Big( x^{|\nu|
-\frac q2 - 2 \alpha_0}e^{q x R} (\widetilde{\gamma} - \log
x)^{q_0 - j_0}
 \Big)\, d x\\ & = \int_{|t|}^\infty x^{-2 s} \Big(
 \frac{|\nu|
-\frac q2 - 2 \alpha_0}{x} + q R -\frac{(q_0 - j_0)}{x (\widetilde{\gamma} - \log x)} \Big)\, d x\\
& =  \Big( |\nu| -\frac q2 - 2 \alpha_0 \Big) \frac{|t|^{-2 s}}{2
s} + q  R\frac{|t|^{- 2 s + 1}}{2 s - 1} +(j_0 - q_0)
\int_{|t|}^\infty x^{-2 s - 1} \frac{1}{\widetilde{\gamma} - \log
x} \, d x .
\end{align*}
From Lemma \ref{lem-logint} we know that
\[
\int_{|t|}^\infty x^{-2 s - 1} \frac{1}{\widetilde{\gamma} - \log
x} \, d x = e^{- 2 s \widetilde{\gamma}} \log s + g(s),
\]
where $g(s)$ is entire such that
\begin{equation} \label{g0formula}
g(0) = \gamma + \log (2 (\log |t| - \widetilde{\gamma})).
\end{equation}
Therefore,
\begin{align*}
\zeta(s,\Ll_{L}) = &\frac{\sin \pi s}{\pi} \Big( |\nu| -\frac q2 -
2 \alpha_0 \Big) \frac{|t|^{- 2 s}}{2 s}+  \frac{\sin \pi
s}{\pi}q R \frac{|t|^{- 2 s + 1}}{2 s - 1}\\
&+\frac{\sin \pi s}{\pi} (j_0 - q_0) e^{- 2 s \widetilde{\gamma}}
\log s  + \frac{\sin \pi s}{\pi} (j_0 - q_0) g(s)
\\
&+\frac{\sin \pi s}{\pi} \int_{|t|}^\infty x^{-2 s} \frac{d}{d x}
\log \Big( 1 + G(x) \Big)\, dx + \frac{1}{2 \pi i} \int_{\gamma_t}
\mu^{-2 s} \frac{F'(\mu)}{F(\mu)}\, d \mu .
\end{align*}
Since
\[
\frac{\sin \pi s}{\pi} (j_0 - q_0) e^{- 2 s \widetilde{\gamma}}
\log s \equiv (j_0 - q_0) s \log s
\]
modulo a function that is $\mathcal{O}(s^2 \log s)$, it follows
that
\begin{align}
\label{zetasLlL} \zeta_0(s,\Ll_{L})  =& \zeta (s,\Ll_{L}) - (j_0 - q_0) s \log s\\
\notag   \equiv & \frac{\sin \pi s}{\pi} \Big( |\nu| -\frac q2 - 2
\alpha_0 \Big) \frac{|t|^{- 2 s}}{2 s} + \frac{\sin \pi s}{\pi}q R
\frac{|t|^{- 2 s + 1}}{2 s - 1} \\
\notag & + \frac{\sin \pi s}{\pi} (j_0 - q_0) g(s)
 +\frac{\sin \pi s}{\pi} \int_{|t|}^\infty
x^{-2 s} \frac{d}{d x} \log \Big( 1 + G(x) \Big)\, dx \notag\\
&\qquad\qquad \qquad\qquad\qquad \qquad \qquad\qquad +\frac{1}{2
\pi i} \int_{\gamma_t} \mu^{-2 s} \frac{F'(\mu)}{F(\mu)}\, d \mu
\notag
\end{align}
modulo a function that is $\mathcal{O}(s^2 \log s)$. The
derivative of the fourth term on the right in \eqref{zetasLlL} is
equal to
\begin{multline} \label{intG}
\cos \pi s \int_{|t|}^\infty x^{-2 s} \frac{d}{d x} \log \Big( 1 +
G(x) \Big)\, dx\\ - \frac{2 \sin \pi s}{\pi} \int_{|t|}^\infty
x^{-2 s} (\log x) \frac{d}{d x} \log \Big( 1 + G(x) \Big)\, dx .
\end{multline}
Since $G(x) = \mathcal{O}\left( \frac{1}{\log x} \right)$  and
$G'(x) = \mathcal{O}\left( \frac{1}{x (\log x)^2}\right)$ as $|x|
\to \infty$ we can put $s = 0$ into the first term in \eqref{intG}
and get
\[
\int_{|t|}^\infty  \frac{d}{d x} \log \Big( 1 + G(x) \Big)\, dx =
- \log \Big( 1 + G(|t|) \Big).
\]
Also using the asymptotics of $G(x)$ and $G'(x)$, we see that the
second term in \eqref{intG} satisfies, for $s \in \R$ with $s \to
0^+$,
\begin{multline*}
\frac{2 \sin \pi s}{\pi} \int_{|t|}^\infty x^{-2 s} (\log x)
\frac{d}{d x} \log \Big( 1 + G(x) \Big)\, dx = \mathcal{O} \left(s
\int_{|t|}^\infty x^{-2 s}  \frac{(\log x)}{x (\log x)^2} \, dx
\right)\\ = \mathcal{O} \left(s \int_{|t|}^\infty x^{-2 s}
\frac{1}{x (\log x)} \, dx \right) = \mathcal{O} \big(s \log s
\big),
\end{multline*}
where we used Lemma \ref{lem-logint} with $c = 0$. In conclusion,
\[
\lim_{s \to 0^+} \frac{d}{d s} \left\{ \frac{\sin \pi s}{\pi}
\int_{|t|}^\infty x^{-2 s} \frac{d}{d x} \log \Big( 1 + G(x)
\Big)\, dx \right\} = - \log \Big( 1 + G(|t|) \Big).
\]
Now, using that
\[
\frac{\sin (\pi s)}{\pi} \Big|_{s = 0} = 0 \ , \ \frac{d}{d s}
\frac{\sin (\pi s)}{\pi} \Big|_{s = 0} = 1 \ , \ \frac{\sin (\pi
s)}{\pi s} \Big|_{s = 0} = 1 \ , \ \frac{d}{d s} \frac{\sin (\pi
s)}{\pi s} \Big|_{s = 0} = 0,
\]
and the formula \eqref{g0formula} for $g(0)$, we can take the
derivatives of the other terms in \eqref{zetasLlL} and set $s = 0$
to conclude that
\begin{align*}
\lim_{s \to 0^+}  \zeta_0 ' (s,\Ll_{L})  =&  - \Big( |\nu| -\frac
q2 - 2 \alpha_0 \Big) \log |t| - q R |t| +(j_0 - q_0) g(0) \\ &-
\log \Big( 1 + G(|t|) \Big)  - \frac{1}{\pi i} \int_{\gamma_t}
\log
\mu \frac{F'(\mu)}{F(\mu)}\, d \mu \\
 = & - \Big( |\nu| -\frac q2 - 2 \alpha_0 \Big) \log |t|
- q R |t| +
(j_0 - q_0) \Big( \gamma + \log (2 (\log |t| - \widetilde{\gamma})) \Big)\\
&  - \log \Big( 1 + G(|t|) \Big)
 - \frac{1}{\pi i} \int_{\gamma_t} \log
\mu \frac{F'(\mu)}{F(\mu)}\, d \mu .
\end{align*}
By definition of $G(x)$, we have
\begin{align*}
\log \Big( 1 + G(|t|) \Big) & = \log \Big( \frac{F(i |t|)}{C
|t|^{|\nu| -\frac q2 - 2 \alpha_0}e^{q |t| R} (\widetilde{\gamma}
- \log
|t|)^{q_0 - j_0} } \Big)\\
& = \log \Big( \frac{F(i |t|)}{C (-1)^{q_0 - j_0}} \Big) - \log
\Big( |t|^{|\nu| -\frac q2 - 2 \alpha_0}e^{q |t| R} (\log |t| -
\widetilde{\gamma})^{q_0 - j_0}   \Big) \\
& = \log \Big( \frac{F(i |t|)}{C (-1)^{q_0 - j_0}} \Big)  - \Big(
|\nu| -\frac q2 - 2 \alpha_0 \Big) \log |t| - q R |t|\\
& \hspace{1em} + (j_0 - q_0) \log (\log |t| - \widetilde{\gamma}).
\end{align*}
Replacing this expression into the preceding expression for
$\lim_{s \to 0^+}  \zeta_0 ' (s,\Ll_{L})$, cancelling appropriate
terms, and using that $F(i |t|) = F(t)$ since $t = i |t|$, we
obtain
\begin{align*}
\lim_{s \to 0^+} \zeta'_0 (s,\Ll_{L}) & = - \log \Big(
\frac{F(t)}{C(-1)^{q_0 - j_0}} \Big) + (j_0 - q_0) \Big( \gamma +
\log 2 \Big) - \frac{1}{\pi i} \! \int_{\gamma_t} \! \! \! \log
\mu \frac{F'(\mu)}{F(\mu)}\, d \mu \\
& = - \log \Big( (-1)^{q_0 - j_0} 2^{q_0 - j_0} e^{(q_0 - j_0)
\gamma} \frac{F(t)}{C} \Big) - \frac{1}{\pi i} \int_{\gamma_t}
\log \mu \frac{F'(\mu)}{F(\mu)}\, d \mu .
\end{align*}
Therefore,
\begin{equation} \label{detreg}
{\det}_{\zeta} (\Ll_L) = (-1)^{q_0 - j_0} 2^{q_0 - j_0} e^{(q_0 -
j_0) \gamma} \frac{F(t)}{C}  \cdot \exp\Big(  \frac{1}{\pi i}
\int_{\gamma_t} \log \mu \frac{F'(\mu)}{F(\mu)}\, d \mu \Big).
\end{equation}
This formula is derived, a priori, when $t$ is on the upper half
part of the imaginary axis. However, the right-hand side is a
\emph{holomorphic} function of $t \in \dD$, where $\dD$ is the set
of complex numbers minus the negative real axis and the zeros of
$F(\mu)$. Therefore \eqref{detreg} holds for all $t \in \dD$. Note
that this equality holds in general even if $\Ll_L$ has a
nontrivial kernel. But to control the factor $\exp ( \frac{1}{\pi
i}\int_{\gamma_t} \cdot\, d\mu)$, we need the condition that $\ker
\Ll_L=\{0\}$. Under this condition, recalling that $\gamma_t$ is
any curve in $\dD$ from $t$ to $-t$, the trick now is to let $t
\to 0$ in \eqref{detreg}, that is, taking $t \to 0$ in $\dD$ from
the upper half plane as shown in Figure
\ref{fig-contstrangedeform}, it follows that
\[
\exp\Big(  \frac{1}{\pi i} \int_{\gamma_t} \log \mu
\frac{F'(\mu)}{F(\mu)}\, d \mu \Big) \to \exp\Big( 0 \Big) = 1.
\]
We also have
\[
F(0) = \det
\begin{pmatrix} \Aa & \Bb \\
\begin{array}{cc} \Id_{q_0} & 0 \\
0 & {\mathbf R}^\nu \end{array} &
\begin{array}{cc} (\log R) \Id_{q_0} & 0 \\
0 & {\mathbf R}^{-\nu} \end{array}
\end{pmatrix}
\]
from Proposition \ref{prop-Fmu}.
\begin{figure}
\centering \includegraphics{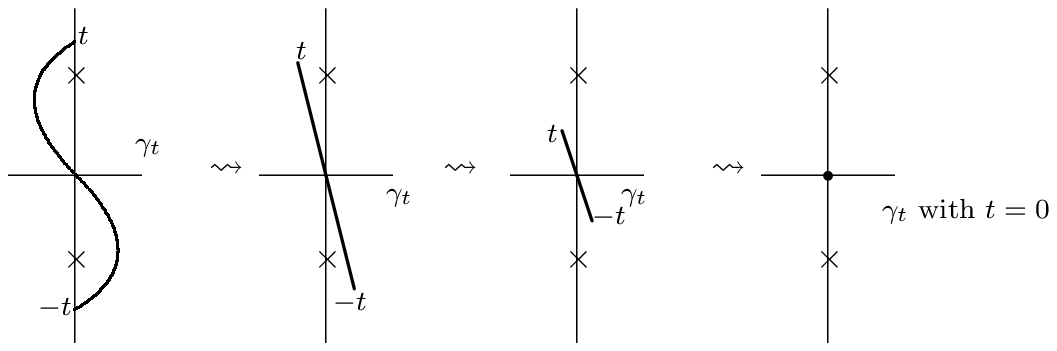}
\caption{The contour $\gamma_t$ as we let $t \to 0$ in $\dD$ from
the upper half plane.} \label{fig-contstrangedeform}
\end{figure}

In conclusion, taking $t \to 0$ on the right side of
\eqref{detreg}, we see that
\[
{\det}_{\zeta} (\Ll_L) =  \frac{(- 2 e^{\gamma})^{q_0 - j_0}}{C}
\det
\begin{pmatrix} \Aa & \Bb \\
\begin{array}{cc} \Id_{q_0} & 0 \\
0 & {\mathbf R}^\nu \end{array} &
\begin{array}{cc} (\log R) \Id_{q_0} & 0 \\
0 & {\mathbf R}^{-\nu} \end{array}
\end{pmatrix}.
\]
Finally, using that $C = a_{j_0 \alpha_0}(2 \pi R)^{-\frac q2}
\prod_{j = 1}^{q_1} 2^{-\nu _{j} } \Gamma (1 - \nu _{j} )$, we get
\begin{multline} \label{detLl}
{\det}_\zeta (\Ll_L ) = \frac{(2 \pi R)^{\frac{q}{2}}}{a_{j_0
\alpha_0}} \prod_{j = 1}^{q_1} \frac {2^{\nu _{j}}}{\Gamma (1 -
\nu _{j} )} (- 2 e^{\gamma})^{q_0 - j_0} \times
\\
{\det
\begin{pmatrix} \Aa & \Bb \\
\begin{array}{cc} \Id_{q_0} & 0 \\
0 & {\mathbf R}^\nu \end{array} &
\begin{array}{cc} (\log R) \Id_{q_0} & 0 \\
0 & {\mathbf R}^{-\nu} \end{array}
\end{pmatrix}
} .
\end{multline}
This completes the proof of Proposition \ref{prop-det}.
\end{proof}

\begin{remark} \em In the case that $\Ll_L$ is not invertible, $F(t)\to F(0)=0$
as $t\to 0$  since $0$ is an eigenvalue of $\Ll_L$. On the other
hand, the left side $\zd (\Ll_L)$ does not depend on $t$. This
means that the factor $\exp(\frac{1}{\pi i}\int_{\gamma_t} \cdot\,
d\mu)$ blows up as $t\to 0$. (Here $\gamma_t$ should not contain
the zero as in Figure \ref{fig-contourdet}.) Therefore, to get the
value of $\zd(\Ll_L)$, we need to know the exact form of the
asymptotics of $F(t)$ and $\exp(\frac{1}{\pi i}\int_{\gamma_t}
\cdot\, d\mu)$ as $t\to 0$.
\end{remark}

Recall that the \emph{Neumann extension} is given by choosing
$\Aa$ and $\Bb$ to be the diagonal matrices with the $q_0 + 1,
\ldots, q$ entries in $\Aa$ equal to $1$ and the $1,\ldots,q_0$
entries in $\Bb$ equal to $1$ with the rest of the entries $0$.
Then the resulting operator $\Ll_\Nn$ has the trivial kernel. This
can be shown as follows: First, by the simple form of $\Aa,\Bb$,
we may assume that $q_0=1, q_1=0$ or $q_0=0, q_1=1$. For the first
case, the solution of $\Ll_L\phi=0$ should have the form $\phi=c_1
r^{\frac12}$ if it exists since the term $r^{\frac12}\log r$
should vanish by the condition of $\Aa,\Bb$ at $r=0$. But, the
Dirichlet condition at $r=R$ implies that $\phi=c_1 r^{\frac12}$
can not be the solution of $\Ll_L$ either. The second case can be
treated in a similar way.  Now we have

\begin{corollary} \label{c:Neumann}
The following equality holds
\[
{\det}_\zeta (\Ll_\Nn ) = (2 \pi R)^{\frac{q}{2}} \prod_{j =
1}^{q_1} \frac {2^{\nu _{j}} \, R^{-\nu_j}}{\Gamma (1 - \nu _{j}
)} .
\]
\end{corollary}
\begin{proof} This proof is just a direct application of the
formula \eqref{detLl}. Observe that for $\Aa$ and $\Bb$ defining
the Neumann extension,
\begin{align*} p(x,y) & := \det
\begin{pmatrix} \Aa & \Bb\\
\begin{array}{cccc}
x\, \Id_{q_0} & 0 & 0 & 0\\
0 & \tau_1\, y^{2 \nu_1} & 0 & 0\\
0 & 0 & \ddots & 0\\
0 & 0 & 0 & \tau_{q_1}\, y^{2 \nu_{q_1}}
\end{array}
&  \Id_q &
\end{pmatrix}\\
& = \det
\begin{pmatrix} 0 & \Id_{q_0}\\
x\, \Id_{q_0} & \Id_{q_0}
\end{pmatrix}
\cdot \det
\begin{pmatrix} \Id_{q_1} & 0\\
\begin{array}{ccc}
\tau_1\, y^{2 \nu_1} & 0 & 0\\
0 & \ddots & 0\\
0 & 0 & \tau_{q_1}\, y^{2 \nu_{q_1}}
\end{array}
&  \Id_{q_1}
\end{pmatrix}\\ & = (-1)^{q_0} x^{q_0} .
\end{align*}
Therefore, $j_0 = q_0$, $\alpha_0 = 0$, and $a_{j_0 \alpha_0} =
(-1)^{q_0}$ for the Neumann extension. In the same way we
simplified $p(x,y)$, we can simplify
\begin{align*} & \det
\begin{pmatrix} \Aa & \Bb \\
\begin{array}{cc} \Id_{q_0} & 0 \\
0 & {\mathbf R}^\nu \end{array} &
\begin{array}{cc} (\log R) \Id_{q_0} & 0 \\
0 & {\mathbf R}^{-\nu} \end{array}
\end{pmatrix} \\
& = \det
\begin{pmatrix} 0 & \Id_{q_0}\\
\Id_{q_0} & (\log R) \Id_{q_0}
\end{pmatrix}
\cdot \det
\begin{pmatrix} \Id_{q_1} & 0\\
{\mathbf R}^{\nu} &  {\mathbf R}^{-\nu}
\end{pmatrix}\\
& = (-1)^{q_0} \prod_{j=1}^{q_1} R^{-\nu_j}.
\end{align*}
Therefore, by \eqref{detLl}, we have
\[
{\det}_\zeta (\Ll_\Nn ) = \frac{(2 \pi
R)^{\frac{q}{2}}}{(-1)^{q_0}} \prod_{j = 1}^{q_1} \frac {2^{\nu
_{j}}}{\Gamma (1 - \nu _{j} )} (- 2 e^{\gamma})^{0} (-1)^{q_0}
\prod_{j=1}^{q_1} R^{-\nu_j} = (2 \pi R)^{\frac{q}{2}} \prod_{j =
1}^{q_1} \frac {2^{\nu _{j}} \, R^{-\nu_j}}{\Gamma (1 - \nu _{j}
)} .
\]
\end{proof}

This corollary agrees with the result in \cite{LMP}. In
particular, for an extension $L$ with $\ker\Ll_L=\{0\}$, we have
\[
\frac{{\det}_\zeta (\Ll_L )}{{\det}_\zeta(\Ll_\Nn)} = \frac{(- 2
e^{\gamma})^{q_0 - j_0}}{a_{j_0 \alpha_0}} \, {\det
\begin{pmatrix} \Aa & \Bb \\
\begin{array}{cc} \Id_{q_0} & 0 \\
0 & {\mathbf R}^{2\nu} \end{array} &
\begin{array}{cc} (\log R) \Id_{q_0} & 0 \\
0 & \Id_{q_1} \end{array}
\end{pmatrix}
} ,
\]
and this formula completes the proof of Theorem \ref{t-det0}.

\section{Special cases of Theorem \ref{t-det0}} \label{sec-special}

In this section we derive various consequences of Theorem
\ref{t-det0}.

\subsection{Row and column conditions}

We begin by proving Theorem \ref{t-det}. Actually, the proof of
Theorem \ref{t-det} follows directly from Theorem \ref{t-det0} and
the following lemma, which computes $a_{j_0 \alpha_0}$ in
\eqref{pxy} explicitly under the row and columns condition of
Theorem \ref{t-det}.

\begin{lemma}
Let $q - r = \mathrm{rank}(\Aa)$ and assume that $\Aa$ has $r$
rows and columns identically zero. Let $i_1,..., i_q$ be a
permutation of the numbers $1,..., q$ such that the rows and
columns $i_1,...,i_{r}$ of $\Aa$ are zero. Choose $j_0 \in
\{0,1,\ldots, r\}$ such that
\[
1 \leq i_1 < i_2 < \cdots < i_{j_0} \leq q_0 < i_{j_0 + 1} <
\cdots < i_r \leq q .
\]
Let $I_{r}$ denote the $q \times q$ matrix which is zero
everywhere except along the diagonal where the entries
$i_1,...,i_{r}$ equal $1$, and let $I_{q-r}$ denote the $q \times
q$ matrix which is zero everywhere except along the diagonal where
the entries $i_{r+1},...,i_q$ equal $1$. Then $\det \left(
\begin{array}{cc}
\Aa & \Bb \\
I_{r} & I_{q - r} \end{array} \right) \ne 0$ and
\[
p(x,y)=a_{j_0,\alpha_0}x^{j_0}y^{2\alpha_0} \big( 1
+\mathcal{O}(|(x,y)|\big)\] where
\[ a_{j_0
\alpha_0} =   \prod_{j=j_0 + 1}^{r} 2^{2 \nu _{i_j}} \frac {\Gamma
(1 + \nu _{i_j} )}{\Gamma (1 - \nu _{i_j} )}\cdot \det \left(
\begin{array}{cc}
\Aa & \Bb \\
I_{r} & I_{q - r} \end{array} \right)
\]
and $\alpha_0 = \nu_{i_{j_0 + 1}} + \nu_{i_{j_0 + 2}} \cdots +
\nu_{i_r}$.

\end{lemma}
\begin{proof} Assume for the moment that $j_0 \geq 1$.
Let $\Aa_1$ denote the matrix $\Aa$ with the $i_1$-th column
removed, let $J_1(x,y)$ denote the matrix
\begin{equation} \label{xytn}
\begin{pmatrix}
x\, \Id_{q_0} & 0 & 0 & 0\\
0 & \tau_1\, y^{2 \nu_1} & 0 & 0\\
0 & 0 & \ddots & 0\\
0 & 0 & 0 & \tau_{q_1}\, y^{2 \nu_{q_1}}
\end{pmatrix}
\end{equation}
with the $i_1$ column and row removed, and finally, let
$\mathcal{C}_1$ denote the $q_1 \times q_1$ identity matrix with
the $i_1$-th row removed. Then expanding the determinant of the
matrix in the definition of $p(x,y)$:
\[
p(x,y) := \det
\begin{pmatrix} \Aa & \Bb\\
\begin{array}{cccc}
x\, \Id_{q_0} & 0 & 0 & 0\\
0 & \tau_1\, y^{2 \nu_1} & 0 & 0\\
0 & 0 & \ddots & 0\\
0 & 0 & 0 & \tau_{q_1}\, y^{2 \nu_{q_1}}
\end{array}
&  \Id_q &
\end{pmatrix},
\]
about the $i_1$-th column, recalling that the $i_1$-th column of
$\Aa$ is zero, we get
\begin{equation} \label{pxyA1}
p(x,y) = \pm x \det
\begin{pmatrix} \Aa_1 & \Bb\\
J_1(x,y) &  \mathcal{C}_1
\end{pmatrix}
\end{equation}
(for an appropriate choice of sign, which happens to equal
$(-1)^{2 i_1 + q}$ in this case). Assume for the moment that $j_0
\geq 2$. Let $\Aa_2$ denote the matrix $\Aa$ with the $i_1$ and
$i_2$ columns removed, let $J_2(x,y)$ denote the matrix
\eqref{xytn} with the $i_1$ and $i_2$ columns and rows removed,
and finally, let $\mathcal{C}_2$ denote the $q_1 \times q_1$
identity matrix with the $i_1$ and $i_2$ rows removed. Then
expanding the determinant of the matrix in \eqref{pxyA1} about the
column containing the zero $i_2$-th column of $\Aa$, we get
\begin{equation} \label{pxyA2}
p(x,y) = \pm x^2 \det
\begin{pmatrix} \Aa_2 & \Bb\\
J_2(x,y) &  \mathcal{C}_2
\end{pmatrix} .
\end{equation}
At this point, we see the general pattern: We expand the
determinant in \eqref{pxyA2} about the column containing the zero
$i_3$-th column of $\Aa$ and then we continue the process of
expanding about each column containing the zero
$i_4,i_5,i_6,\ldots, i_r$ columns of $\Aa$. At the end, we arrive
at
\begin{equation} \label{pxyA3}
p(x,y) = \pm \tilde{\tau} \, x^{j_0} y^{2 \tilde{\nu}} \det
\begin{pmatrix} \Aa_r & \Bb\\
J_r(x,y) &  \mathcal{C}_r
\end{pmatrix},
\end{equation}
where $\Aa_r$ denotes the matrix $\Aa$ with the $i_1,\ldots, i_r$
columns removed, $J_r(x,y)$ denotes the matrix \eqref{xytn} with
the $i_1,\ldots, i_r$ columns and rows removed, and
$\mathcal{C}_r$ denotes the $q \times q$ identity matrix with the
$i_1,\ldots,i_r$ rows removed.

Now observe that
\[
\pm \det
\begin{pmatrix} \Aa_r & \Bb\\
J_r(0,0) &  \mathcal{C}_r \end{pmatrix} = \pm \det
\begin{pmatrix} \Aa_r & \Bb\\
0 &  \mathcal{C}_r \end{pmatrix} = \det \left(
\begin{array}{cc}
\Aa & \Bb \\
I_{r} & I_{q - r} \end{array} \right);
\]
indeed, the first equality is obvious because $J_r(0,0)$ is the
zero matrix while the second equality can be easily verified by
expanding the determinant $\det \left(
\begin{array}{cc}
\Aa & \Bb \\
I_{r} & I_{q - r} \end{array} \right)$ about the zero $i_1,i_2,
\ldots, i_r$ columns of $\Aa$ just as we did in the previous
paragraph.

It remains to prove that $\det \left(
\begin{array}{cc}
\Aa & \Bb \\
I_{r} & I_{q - r} \end{array} \right)  \ne 0$. To see this, recall
that the $i_1,\ldots, i_r$ rows of $\Aa$ are identically zero.
This implies that, since the rank of $\Aa$ is $q - r$, the rows of
$\Aa$ complementary to $i_1,\ldots, i_r$, namely the
$i_{r+1},\ldots, i_q$ rows where we use the notation as in the
statement of this lemma, are linearly independent. Therefore,
since the matrix $\begin{pmatrix} \Aa & \Bb
\end{pmatrix}$ has rank $q$, the $i_1,\ldots,i_r$ rows of
$\Aa$ are identically zero, and the $i_{r+1},\ldots,i_q$ rows of
$\Aa$ are linearly independent, it follows that the
$i_1,\ldots,i_{r}$ rows of $\Bb$ are linearly independent and
these rows, together with the $i_{r+1},\ldots,i_{q}$ rows of $\Aa$
span all of $\C^q$. Now recall that the $i_1,\ldots,i_r$ columns
of $\Aa$ are identically zero; in particular, the span of the
$i_{r+1},\ldots,i_{q}$ rows of $\Aa$ does not contain any
$e_{i_1},\ldots, e_{i_r}$, where $e_j$ denote the unit vector in
$\C^q$ with $j$-th slot equal to $1$ and $0$'s elsewhere. It
follows that the span of the $i_{r+1},\ldots,i_{q}$ rows of $\Aa$
(which are linearly independent) is contained in the span of
$e_{i_{r+1}},\ldots, e_{i_{q}}$. Therefore, by the property of
dimension,
\begin{equation} \label{fact1}
\text{the span of the $i_{r+1},\ldots,i_{q}$ rows of $\Aa$ $=$ the
span of $e_{i_{r+1}},\ldots, e_{i_q}$.}
\end{equation}
Hence, as the $i_1,\ldots,i_{r}$ rows of $\Bb$ plus the
$i_{r+1},\ldots,i_{q}$ rows of $\Aa$ span all of $\C^q$, it
follows that
\begin{equation} \label{fact2}
\text{the span of the $i_1,\ldots,i_{r}$ rows of $\Bb$ $=$ the
span of $e_{i_1},\ldots, e_{i_r}$}.
\end{equation}
We are now ready to prove our lemma. The nonzero rows of
\[
\begin{pmatrix} \Aa \\ I_r \end{pmatrix}
\]
are linearly independent by \eqref{fact1}. The rows in the matrix
\[
\begin{pmatrix} \Bb \\ I_{q - r} \end{pmatrix}
\]
that are complementary to the nonzero rows of $\begin{pmatrix} \Aa \\
I_r \end{pmatrix}$ are therefore linearly independent by
\eqref{fact2}. It follows that the matrix $\left(
\begin{array}{cc}
\Aa & \Bb \\
I_{r} & I_{q - r} \end{array} \right)$ has full rank, which is
equivalent to $\det \left(
\begin{array}{cc}
\Aa & \Bb \\
I_{r} & I_{q - r} \end{array} \right)  \ne 0$. Now the formula of
$a_{j_0\alpha_0}$ follows from \eqref{pxy} and \eqref{pxyA3}. This
completes the proof.
\end{proof}

\subsection{Decomposable Lagrangians}

Because the $-\frac 14$ eigenvalues and the eigenvalues in
$(-\frac14,\frac34)$ of $A_\Gamma$ result in rather different
analytic properties, it is natural to separate these eigenvalues.
With this discussion in mind, we shall call a Lagrangian subspace
$L \subset V$ \emph{decomposable} if $L = L_0 \oplus L_1$ where
$L_0$ is a Lagrangian subspace of $\bigoplus_{\la_\ell = -\frac14}
E_\ell\oplus E_\ell$ and $L_1$ is a Lagrangian subspace of
$\bigoplus_{-\frac14<\la_\ell<\frac34} E_\ell\oplus E_\ell$. As
described in Proposition \ref{p:relation}, the Lagrangian subspace
$L_0$ is determined by two $q_0 \times q_0$ matrices $\Aa_0$,
$\Bb_0$ where $q_0 = \dim L_0$, that is, the multiplicity of the
eigenvalues $\la_\ell = -\frac14$. Similarly, the Lagrangian
subspace $L_1$ is determined by two $q_1 \times q_1$ matrices
$\Aa_1$, $\Bb_1$ where $q_1 = \dim L_1$, that is, the multiplicity
of the eigenvalues $\la_\ell$ with $-\frac14<\la_\ell<\frac34$.
Thus, the function $p(x,y)$ in \eqref{polyp} takes the form
\begin{align*} p(x,y) & := \det
\begin{pmatrix}
\begin{array}{cc}
\Aa_0  & 0\\
0  & \Aa_1
\end{array} &
\begin{array}{cc}
\Bb_0 & 0\\
0 & \Bb_1
\end{array}
\\
\begin{array}{cccc}
x\, \Id_{q_0} & 0 & 0 & 0\\
0 & \tau_1\, y^{2 \nu_1} & 0 & 0\\
0 & 0 & \ddots & 0\\
0 & 0 & 0 & \tau_{q_1}\, y^{2 \nu_{q_1}}
\end{array}
&  \Id_q &
\end{pmatrix}\\
& = \det
\begin{pmatrix} \Aa_0 & \Bb_0\\
x\, \Id_{q_0} & \Id_{q_0}
\end{pmatrix}
\cdot \det
\begin{pmatrix} \Aa_1 & \Bb_1\\
\begin{array}{ccc}
\tau_1\, y^{2 \nu_1} & 0 & 0\\
0 & \ddots & 0\\
0 & 0 & \tau_{q_1}\, y^{2 \nu_{q_1}}
\end{array}
&  \Id_{q_1}
\end{pmatrix}\\
& = : p_0(x) \cdot p_1(y),
\end{align*}
where $p_0$ and $p_1$ are the corresponding determinants in the
second line. Expanding the determinants, we can write
\begin{equation} \label{p01}
p_0(x) = \sum a_j \, x^j \qquad \text{and}\qquad p_1(y) = \sum
b_\alpha \, y^{2 \alpha}.
\end{equation}
The next theorem follows immediately from Proposition
\ref{prop-det} and Theorem \ref{t-det0}.

\begin{theorem}\label{t-det1}
For a decomposable Lagrangian $L \subset \C^{2q}$ such that
$\ker\Ll_L=\{0\}$, we have
\begin{multline} \label{detL}
{\det}_\zeta (\Ll_L ) = \frac{(2 \pi R)^{\frac{q}{2}}}{a_{j_0}
b_{\alpha_0}} \prod_{j = 1}^{q_1} \frac {2^{\nu _{j}}}{\Gamma (1 -
\nu _{j} )} (- 2 e^{\gamma})^{q_0 - j_0} \times
\\
{\det
\begin{pmatrix} \Aa_0 & \Bb_0 \\
\Id_{q_0} &  (\log R) \Id_{q_0}
\end{pmatrix}
\det\left(
\begin{array}{cc} \Aa_1 & \Bb_1 \\
{\mathbf R}^{ \nu} & {\mathbf R}^{-\nu} \end{array} \right)} ,
\end{multline}
where $a_{j_0}$ and $b_{\alpha_0}$ are the coefficients in
\eqref{p01} corresponding to the smallest $j$ and $\alpha$ with a
nonzero coefficient in $p_0(x)$ and $p_1(y)$, respectively. In
particular, for the generalized cone we have
\[
\frac{{\det}_\zeta (\Delta_L )}{{\det}_\zeta(\Delta_\Nn)} =
\frac{(- 2 e^{\gamma})^{q_0 - j_0}}{a_{j_0} b_{\alpha_0}} \, {\det
\begin{pmatrix} \Aa_0 & \Bb_0 \\
\Id_{q_0} &  (\log R) \Id_{q_0}
\end{pmatrix}
\det\left(
\begin{array}{cc} \Aa_1 & \Bb_1 \\
{\mathbf R}^{2 \nu} & \Id_{q_1} \end{array} \right)}  .
\]
\end{theorem}

\subsection{The one-dimensional case}

Consider now the one-dimensional operator
\[
\Ll := - \frac{d^2}{d r^2} + \frac{1}{r^2} \la \quad \text{over\ \
$[0,R]$,\ \ where $- \frac14 \leq \la < \frac34$.}
\]
In this one-dimensional case, Lagrangians are given by two $1
\times 1$ matrices (numbers) $\Aa = \alpha$ and $\Bb = \beta$
where $\alpha \overline{\beta} \in \R$. One can check that (see
e.g.\ \cite[prop.\ 3.7]{KLP} that we can take $\alpha ,\beta
\in\R$ with $\alpha^2 + \beta^2 = 1$. We shall compute
${\det}_\zeta (\Ll_L)$ using Theorem \ref{t-det1} under the
assumption $\ker\Ll_L=\{0\}$. Assume that $\la = - \frac14$. Then
\[
p_0(x) = \det
\begin{pmatrix} \alpha & \beta \\ x & 1
\end{pmatrix} = \alpha - \beta \, x ,
\]
which implies that $j_0 = 0$ and $a_{j_0} = \alpha$ if $\alpha \ne
0$ and $j_0 = 1$ and $a_{j_0} = - \beta$ if $\alpha = 0$, and by
\eqref{detL}, we have
\begin{align*}
{\det}_\zeta (\Ll_L ) & = \frac{(2 \pi R)^{\frac{1}{2}}}{a_{j_0}}
(- 2 e^{\gamma})^{1 - j_0} \det
\begin{pmatrix} \alpha & \beta \\1 & (\log R)
\end{pmatrix} \\ & = \frac{\sqrt{2 \pi R}}{a_{j_0}}
(- 2 e^{\gamma})^{1 - j_0} \Big( \alpha \log R - \beta \Big).
\end{align*}
In conclusion, we see that in the case $\la = - \frac14$, we have
\[
{\det}_\zeta (\Ll_L ) = \begin{cases} 2  \sqrt{2 \pi R}\,
e^{\gamma} \Big( \frac{\beta}{\alpha} - \log R  \Big) & \text{if $\alpha \ne 0$}\\
\sqrt{2 \pi R} & \text{if $\alpha = 0$.} \end{cases}
\]
Assume now that $- \frac14 < \la < \frac34$. Then with $\nu :=
\sqrt{\la + \frac14}$ and $\tau = 2^{2\nu} \frac{ \Gamma(1 +
\nu)}{ \Gamma(1 - \nu)}$, we have
\[
p_1(y) = \det
\begin{pmatrix} \alpha & \beta \\ \tau \, y^{2 \nu} & 1
\end{pmatrix} = \alpha - \beta \, \tau \, y^{2 \nu} ,
\]
which implies that $\alpha_0 =0$ and $b_{\alpha_0} = \alpha$ if
$\alpha \ne 0$ and $\alpha_0 = 2\nu$ and $b_{\alpha_0} = - \beta
\, \tau$ if $\alpha = 0$, and by \eqref{detL}, we have
\[
{\det}_\zeta (\Ll_L ) = \frac{(2 \pi
R)^{\frac{1}{2}}}{b_{\alpha_0}} \frac {2^{\nu }}{\Gamma (1 - \nu
)} \det
\begin{pmatrix} \alpha & \beta \\ R^{ \nu} & R^{-\nu}
\end{pmatrix} = \frac{\sqrt{2 \pi R}}{b_{\alpha_0}}\frac {2^{\nu }}{\Gamma (1 - \nu
)} \Big( \alpha R^{-\nu} - \beta R^{\nu}\Big).
\]
In conclusion, we see that in the case $- \frac14 < \la <
\frac34$, we have
\[
{\det}_\zeta (\Ll_L ) = \begin{cases}  {2^{\nu +1/2}}\, \sqrt{ \pi
R}\  {\Gamma (1 - \nu )^{-1}} \, \Big( R^{-\nu}
- \frac{\beta}{\alpha} R^{\nu} \Big) & \text{if $\alpha \ne 0$}\\
{2^{-\nu +1/2}}\,{\sqrt{ \pi R}}\ {\Gamma (1 + \nu )^{-1}}\,
R^{\nu} & \text{if $\alpha = 0$.}
\end{cases}
\]

\section{Conclusions and final remarks}
In this article we have considered zeta functions and zeta
regularized determinants for arbitrary self-adjoint extensions of
Laplace-type operators over conic manifolds. In general, the zeta
function will have a logarithmic branch point as well as a simple
pole at $s=0$. In order to get a well-defined notion of a
determinant we propose to use the natural prescription
(\ref{defdet}). Within this prescription, Theorem \ref{t-det0} is
the central theorem proven in this article. It gives a closed form
for the determinant of the Laplacian over the cone associated with
an arbitrary self-adjoint extension. As we have seen, it is easily
applied to particular cases and known results have been easily
reproduced.

For convenience we have chosen to work with Dirichlet boundary
conditions at $r=R$, emphasizing the role of the self-adjoint
extension for the analytic structure of the zeta function and for
the determinant. Equally well other boundary conditions at $r=R$
can be considered along the same lines.

\section*{Acknowledgements}
KK was supported in part by funds from the Baylor University
Research Committee, by the Baylor University Summer Sabbatical
Program and by the Max-Planck-Institute for Mathematics in the
Sciences (Leipzig, Germany).

\bibliographystyle{amsplain}
\def\cprime{$'$} \def\polhk#1{\setbox0=\hbox{#1}{\ooalign{\hidewidth
  \lower1.5ex\hbox{`}\hidewidth\crcr\unhbox0}}}
\providecommand{\bysame}{\leavevmode\hbox
to3em{\hrulefill}\thinspace}
\providecommand{\MR}{\relax\ifhmode\unskip\space\fi MR }
\providecommand{\MRhref}[2]{%
  \href{http://www.ams.org/mathscinet-getitem?mr=#1}{#2}
} \providecommand{\href}[2]{#2}

\end{document}